\documentclass[12pt]{article}
\usepackage{amsmath,amsthm,amsfonts}
\usepackage{cite,citesort}
\def \BE {\begin{equation}}
\def \EE {\end{equation}}
\def \BEA {\begin{eqnarray}}
\def \EEA {\end{eqnarray}}
\def \CR {\nonumber \\}
\def \nn {\nonumber}

\def \p {\partial}
\def \ddt {\frac{\partial}{\partial t}}

\def \nabx {\nabla_x}
\def \wkx {\omega_{kx}}
\def \muev {\mu_{ev}}
\def \muod {\mu_{od}}
\def \alev {\alpha_{ev}}
\def \alod {\alpha_{od}}
\def \wev {\omega_{ev}}
\def \wod {\omega_{od}}
\def \al {\alpha}
\def \W {\Omega}
\def \w {\omega}
\def \i {i}
\def \at {\tilde{a}}
\def \ac {\check{a}}

\def \aee {{\textbf{m}}/2}
\def \aeb {\textbf{m}}
\def \ep {\varepsilon}
\def \eps {\varepsilon^*}
\def \spi {\sqrt{\pi}}

\def \psih {\hat{\psi}}
\def \psit {\tilde{\psi}}
\def \psic {\check{\psi}}
\def \hf {\hat{f}}
\def \cn {\check{n}}
\def \cnw {{n}^{\rm W}}
\def \k {\textbf{k}}
\def \q {\textbf{q}}
\def \x {\textbf{x}}
\def \s {\textbf{s}}

\def \p {\textbf{p}}

\def \l {\textbf{l}}
\def \m {\textbf{m}}
\def \s {\textbf{s}}
\def \nkx {(\nabla_\k\cdot\nabla_\x)}
\def \a {\alpha}
\def \b {\beta}
\def \g {\gamma}
\def \alk {\a_\k}
\def \bmk {\b_{-\k}}
\def \gmk {\g_{-\k}}
\def \ak {\a_\k}
\def \bk {\b_\k}
\def \gk {\g_\k}

\newtheorem*{lem}{Lemma}
\newtheorem*{thm}{Theorem}

\begin{document}

\title{Canonical  Hamiltonians for waves in inhomogeneous media}

\author{Boris Gershgorin$^a$, Yuri V. Lvov$^b$, and Sergey Nazarenko$^c$, \\ \ \\
$^a$ \textit{Courant Institute, New York University, New York, NY 
10012, USA}\\
$^b$ \textit{Department of Mathematical Sciences, 
Rensselaer Polytechnic Institute,} \\ \textit{ Troy, NY 12180, USA}\\
$^c$ \textit{Mathematics Institute, University of Warwick, Coventry CV4 7AL, UK}\\
}
\maketitle

\small

\begin{abstract}
We obtain a canonical form of a quadratic Hamiltonian for linear waves
in a weakly inhomogeneous medium.  This is achieved by using the WKB
representation of wave packets.  The canonical form of the Hamiltonian
is obtained via the series of canonical Bogolyubov-type and
near-identical transformations.  Various examples of the application
illustrating the main features of our approach are presented.  The
knowledge of the Hamiltonian structure for linear wave systems
provides a basis for developing a theory of weakly nonlinear random
waves in inhomogeneous media generalizing the theory of homogeneous
wave turbulence.
\end{abstract} 

\section{Introduction}
In order to analyze the behavior of a general nonlinear system, the
first necessary step is to analyze the linear system.  The classical
examples are finite dimensional Hamiltonian systems with coupled
degrees of freedom. The dynamical behavior of such systems for small
excitations, is determined by the form of the quadratic part of their
Hamiltonians, which correspond to the linear dynamics.  A detailed
classification of the canonical normal forms for such Hamiltonians was
given by Galin and it was summarized in one of the appendices of
Arnold's book~\cite{Arnold}.
Another class of examples of nonlinear systems, which behavior
crucially depends on the linear part, is weakly interacting dispersive
waves in continuous media that are studied in Wave Turbulence (WT).
These systems often have analogs among discrete systems, e.g. a chain
of coupled oscillators, and correspondingly their quadratic
Hamiltonians have analogs among Galin-Arnold canonical forms.
Examples where WT has important applications are surface gravity
waves~\cite{Zakharov_grav}, $\beta$-plane turbulence in oceans and
atmospheres~\cite{monin_piterbarg,Balk_naz_zakh}, internal waves in
the ocean~\cite{Lvov_Tabak}, weak magnetohydrodynamic (MHD) turbulence
\cite{galtier_etal_2000}, Bose-Einstein condensate \cite{BEC,BEC_LN}
and plasmas \cite{Plasma}.  Traditionally, WT theory is applied to
homogeneous systems.  The quadratic interaction Hamiltonian with linear interactions describing
such homogeneous systems is given by~\cite{ZLF}
\BEA H_2=\int\left(
A_\k|a_\k|^2d\k+\frac{1}{2}(B_\k a_\k
a_{-\k}+B_\k^*a_\k^*a_{-\k}^*)\right)d\k,\label{Ham} \EEA
where $a_\k$ is a complex-valued field, the bold face denotes a $d$
dimensional vector in $\mathbb{R}^d$.
  The linear canonical transformation to the new variables
$b_\k$ that was first used by Bogolyubov in 1958 for the system of
fermions, brings this Hamiltonian to its normal form 
\BEA H_2=\int\w_\k|b_\k|^2d\k,\label{intr_canon} \EEA
where $\w_\k$ is a
linear dispersive relationship.  
The formalism of WT significantly enhanced our understanding of spectral energy transfer in ocean, atmosphere,
plasma, other system of nonlinear waves~\cite{ZLF}. 
Wave turbulence deals with weakly nonlinear waves with quasi-random phases.
Using WT, one can derive a kinetic equation for the wave spectrum, which evolves due to resonant wave interactions.

In order for the resonant energy transfers to occur, certain resonance
conditions need to be satisfied. In particular, for the systems
dominated by three wave interactions, such as internal waves in the
ocean~\cite{Lvov_Tabak} or capillary waves~\cite{Capil}, these conditions are given by
\BEA
\w_\k&=&\w_{\k_1}+\w_{\k_2},\nonumber\\ \k&=&\k_1+\k_2.\nonumber \\
\label{3wave}
\EEA
Similar four-wave resonant conditions should be satisfied for the resonant interactions in the four-wave systems.

Often, WT is not spatially homogeneous and its statistical properties
vary in space due to a trapping potential, inhomogeneous background
density or an inhomogeneous velocity field.  Examples of such systems
are the Bose-Einstein Condensate (BEC) in the presence of a trapping
potential~\cite{BEC_LN} and an interaction of the long "aged" gravity
waves with a swell~\cite{Aged}.  In general, the idea is to consider a
small amplitude high frequency perturbation of a large-scale solution
of the dynamic equation (e.g. the condensate).  The effects of this
coordinate dependent background solution can most easily be understood
using a wave-packet formalism.  This formalism was first used to
approximate the Schr\"{o}dinger's wave function by a
quasi-monochromatic wave by Wentzel~\cite{W}, Kramers~\cite{K}, and
Brillouin~\cite{B}.  Their initials give the term WKB
approximation. The WKB approximation is applicable if the wavepacket
wavelength $l$ is much shorter than the characteristic wavelength $L$
of a large scale solution
\BEA \epsilon=\frac{l}{L}\ll 1.\nonumber \EEA
The essence of WKB approach is that the wave numbers, characterizing
the wavepacket are the functions of coordinates. This is due to the
distortion of the wave packets by the media, leading to a spatial
wavenumber dependence.  Such spatial wave number dependence may have a dramatic effect on
nonlinear resonant wave interactions. Indeed, the resonant conditions
(\ref{3wave}) may now be satisfied only in a finite part of a domain,
or for a particular wave packet, only for a finite time.

The goal of this paper is to use the spatially dependent WKB wave
packets to find a canonical form for the quadratic Hamiltonian for the
inhomogeneous systems.  This problem can be considered as an extension
of certain (oscillatory) members from the Galin-Arnold classification
of quadratic forms onto the infinite-dimensional and continuous space.
The physical motivation for such a formulation is that, since the
Hamiltonian description is natural for WT in homogeneous systems, it
lays down a necessary framework for generalization onto the
inhomogeneous media.

To begin, we write down the general Hamiltonian for the system of linear waves propagating in the inhomogeneous background. 
We will show in Section~\ref{Motivation} that such general Hamiltonian for the variable $a_\q$ is given by the following quadratic form:
\BEA
H=\int \left(A(\q,\q_1)a_\q a^*_{\q_1}+\frac{1}{2}
\Big(B(\q,\q_1)a_\q a_{-\q_1}+B^*(\q,\q_1)a^*_\q a^*_{-\q_1}\Big)\right)d\q d\q_1.
\label{intr_genHamiltonian}
\EEA
The main result of the present paper is that this general
Hamiltonian~(\ref{intr_genHamiltonian}) can be transformed to the
following canonical form
\BEA
H=\int c_{\k\x}[\w_{\k\x}-\x\cdot\nabla_\x\w_{\k\x}+\i\{\w_{\k\x},\cdot\}]c_{\k\x}^*d\k d\x.\label{canon_inhom}
\EEA
Here $\w_{\k\x}$ and $c_{\k\x}$ are the new position-dependent
dispersion relationship and normal field-variable, correspondingly,
and a Poisson bracket is defined by
\BEA
\{f,g\}=\nabla_\k f\cdot\nabla_\x g-\nabla_\k g\cdot\nabla_\x f.
\nn\EEA

This novel canonical Hamiltonian is a generalization of the
Hamiltonian (\ref{intr_canon}) for the inhomogeneous systems.  The
approach used in the paper can be viewed as a generalization of the
Bogolyubov transformation, which diagonalizes Hamiltonian~(\ref{Ham})
to the form~(\ref{intr_canon}) via canonical transformation.
Similarly, Hamiltonian~(\ref{intr_genHamiltonian}) can be transformed
into the canonical form (\ref{canon_inhom}), however, now using
near-canonical transformations.  In the
Hamiltonian~(\ref{canon_inhom}), the second and the third terms in the
brackets correspond to the higher order corrections to the dispersion
relation due to the inhomogeneity.  We prove that the
Hamiltonian~(\ref{intr_genHamiltonian}) can be transformed into a
canonical form~(\ref{canon_inhom}) in the case when the heterogeneity
is weak. Formally, the requirement of weak inhomogeneity means that
the coefficients $A(\q,\q_1)$ and $B(\q,\q_1)$ are strongly peaked at
$\q-\q_1=0$, i.e. $A(\q,\q_1)=0$ for $|\q-\q_1|>\ep$ for some small
$\ep$.  Based on this requirement, we derive below the re-normalized
dispersion relationship and the transformation formulas from $a_\k$ to
$c_{\k\x}$ accurate up to the first order in $\ep$.  It turns out that
just Bogolyubov's transformation is not enough in this case.  The
phase coordinate systems should also be perturbed by a near-identity
transformation in addition to the Bogolyubov's rotation.  Then, the
Hamiltonian becomes diagonal up to the first order in $\epsilon$.

From the novel canonical form of the Hamiltonian given by
Eq.~(\ref{canon_inhom}), the traditional radiative action balance
equation can easily be obtained:
\BEA
\frac{\partial n_{\k\x}}{\partial t}+\nabla_\k\w_{\k\x}\nabla_{\x}n_{\k\x}-\nabla_\x\w_{\k\x}\nabla_{\k}n_{\k\x}=0,\nonumber %
\EEA
or, shorter,
\BEA
\frac{\partial n_{\k\x}}{\partial t}+\{\w_{\k\x},n_{\k\x}\}=0,
\label{kinetic}
\EEA
where $n_{\k\x}$ denotes the ensemble average of the squared amplitude
of the wave, i.e., $n_{\k\x}\equiv\langle |c_{\k\x}|^2\rangle$.
Equation~(\ref{kinetic}) is now a standard equation which is used in
statistical modeling of wave systems~\cite{janssen}. It is a wave
analog of the Liouville's theorem, or the continuity equations for
distribution function of statistical
mechanics~\cite{LandauStat,LandauKin}.  In wave systems there are wave
quasi-particles instead of particles, following different rays instead
of individual particle trajectories. Wave action distribution function
moves in the multidimensional wave-number-coordinate  phase space.

The paper is organized as follows.  In Section~\ref{Motivation}, we
give simple and instructive examples that motivate the study of
inhomogeneous WKB systems.  In Section~\ref{Preliminaries}, we
introduce the window transforms and other formulas that will be
extensively used later.  In Section~\ref{Lemma}, we discuss the case
of a nearly-diagonal Hamiltonian.  We show how it can be transformed
to a canonical form~(\ref{intr_canon}) and provide a couple of
representative examples.  In Section~\ref{Theorem}, we study the
Hamiltonian in a general form~(\ref{intr_genHamiltonian}).  We present
the series of near-canonical and near-identical transformations that
bring the Hamiltonian~(\ref{intr_genHamiltonian}) to the
form~(\ref{canon_inhom}).  We also demonstrate the application of this
approach to the nonlinear Schr\"odinger equation with the condensate.

\section{Motivation\label{Motivation}}
In this Section, we motivate our study of inhomogeneous systems by considering an example of the interaction of short waves on the background of a long wave.
We describe the cases of both three-wave and four-wave Hamiltonian systems. 

\paragraph*{Three-wave case.}
The quadratic Hamiltonian of a three-wave system with small scale
perturbations on the background of the large scale excitations is
given by Eq.~(\ref{intr_genHamiltonian}). 
In order to show that, we start with a standard three-wave
Hamiltonian~\cite{ZLF}:
\BEA H_3=\int\W_\k|a_\k|^2d\k+\frac{1}{2}\int
\Big(V^\k_{\l\m}a_\k^*a_\l a_\m+c.c.\Big)\delta^\k_{\l\m} d\k d\l
d\m,\label{H3} \EEA
where $V$ is an interaction coefficient. 
Then, the equations of motion  for the variable $a_\k$ assume a standard form
\BEA
\i\dot{a}_\k=\frac{\delta H_3}{\delta a_k^*} =
\W_{\k}a_{\k}+\int \Big(\frac{1}{2} V^\k_{\l\m}a_\l a_\m \delta^\k_{\l\m}+\left(V^\l_{\k\m}\right)^*a_\l a_\m^*\delta^\l_{\k\m}\Big) d\l d\m.\label{eqnofmotion}
\EEA
Suppose that a large-scale solution of (\ref{eqnofmotion}) is given by $C_\k$.
We consider a perturbed solution $a_\k=C_\k+c_\k$ where $c_\k$ is a small-scale perturbation of $C_k$.  
Equation of motion for $c_\k$ attains the following form:
\BEA
\i\dot{c}_\k=\W_{\k}c_{\k}+\int \left[ \frac{1}{2} V^\k_{\l\m}(C_\l c_\m+C_\m c_\l+c_\l c_\m)\delta^\k_{\l\m}+
 \left(V^\l_{\k\m}\right)^*(C_\l c_\m^*+C_\m^*c_\l+c_\l c_\m^*)\delta^\l_{\k\m} \right] d\l d\m.\label{eq_motion1}
\nn\EEA
Now, we use the fact that $C_\k$ is a known exact solution for Eq.~(\ref{eqnofmotion}) to obtain
\BEA
\i\dot{c}_\k=\int A(\k,\l)c_\l d\l+\frac{1}{2}\int B(\k,\l)c_{-\l}^* d\l+
\int \left[ \frac{1}{2} V^\k_{\l\m}c_\l c_\m \delta^\k_{\l\m}+\left(V^\l_{\k\m}\right)^*c_\l c_\m^* \delta^\l_{\k\m} \right]
d\l d\m,\label{H3_eq}
\EEA
where
\BEA
A(\k,\l)&=&\W_\l\delta_\l^\k+V^\k_{\l,\k-\l}C_{\k-\l}+\left( V^\l_{\k,\l-\k}\right)^*C^*_{\l-\k},\\
B(\k,\l)&=&2V^{\k-\l}_{\k,-\l}C_{\k-\l}.
\nn\EEA
Equation~(\ref{H3_eq}) corresponds to the following Hamiltonian
\BEA
H=\int A(\k,\l)c_\l c_\k^*d\l d\k+\frac{1}{2}\int \left[ B(\k,\l)c_{-\l}^*c_\k^*+c.c.\right]d\k d\l+
\frac{1}{2}\int [V^\k_{\l\m}c_\k^*c_\l c_\m \delta^\k_{\l\m}+c.c.]d\k d\l d\m.\label{H3_hami}
\EEA
This appears to be a standard form of the Hamiltonian for the wave
system dominated by three wave interactions in the inhomogeneous
media.  Quadratic in $c_\k$ part of this Hamiltonian is given by
(\ref{intr_genHamiltonian}), while cubic part of this Hamiltonian is a
standard three-wave interaction Hamiltonian.
\paragraph*{Four-wave case.}
Four-wave systems are similar to three-wave systems when small scale
perturbations on the background of the large scale excitations are
considered. Indeed, we show below that the quadratic part of a
four-wave Hamiltonian of small scale perturbation has the form
(\ref{intr_genHamiltonian}).  We start from a standard four-wave
Hamiltonian~\cite{ZLF}:
\BEA
H_4=\int\W_\k|a_\k|^2d\k+\frac{1}{2}\int T^{\k\l}_{\m\s}a_\k^*a_\l^* a_\m a_\s\delta^{\k\l}_{\m\s} d\k d\l d\m d\s,\label{H4}
\EEA
where $T$ is an interaction coefficient.  The corresponding equation
of motion takes the form
\BEA
\i\dot{a}_\k=\W_{\k}a_{\k}+\int T^{\k\l}_{\m\s}a_\l^* a_\m a_\s \delta^{\k\l}_{\m\s} d\l d\m d\s.
\label{fourwave}
\EEA
We then consider a perturbed solution $a_\k=C_\k+c_\k$ where $c_\k$ is
a small-scale perturbation.  Assuming that $C_\k$ is an exact solution
to the equation of motion with Hamiltonian (\ref{fourwave}), we obtain
the following equation of motion for $c_\k$
\BEA
\i\dot{c}_\k&=&
\W_{\k}c_{\k}+\int T^{\k\l}_{\m\s}(2C_\l^* C_\m c_\s+c_\l^* C_\m C_\s)\delta^{\k\l}_{\m\s}d\l d\m d\s+
\CR &&
                           \int T^{\k\l}_{\m\s}(C_\l^* c_\m c_\s+2c_\l^* c_\m C_\s)\delta^{\k\l}_{\m\s}d\l d\m d\s
                           +\int T^{\k\l}_{\m\s} c_\l^* c_\m c_\s\delta^{\k\l}_{\m\s}d\l d\m d\s.
\nn\EEA
Since $C_\k$ is a known large scale solution, we obtain 
\BEA
\i\dot{c}_\k&=&\int A(\k,\s)c_\s d\s+\frac{1}{2}\int B(\k,\s)c_{-\s}^* d\s
\CR 
&+&\int \left[ \frac{1}{2} W^{\k}_{\l\m}c_\l c_\m+ (W^{\l}_{\k\m})^* c_\l c_\m^* \right] d\l d\m
\CR
  & +&\int T^{\k\l}_{\m\s} c_\l^* c_\m c_\s\delta^{\k\l}_{\m\s}d\l d\m d\s,\label{H4_eq}
\EEA
where we defined the kernels $A(\k,\s)$ and $B(\k,\s)$ as
\BEA
A(\k,\s)&=&\W_\s\delta_\s^\k+2\int T^{\k\l}_{\m\s}C^*_\l C_\m \delta^{\k\l}_{\m\s}d\l d\m\\
B(\k,\s)&=&2\int T^{\k,-\s}_{\m,\l}C_\m C_\l \delta^{\k,-\s}_{\m,\l}d\l d\m,
\nn\EEA
and 
\BEA W^{\k}_{\l\m} = 2 \int T^{\k\s}_{\m\l} C_\s^*
\delta^{\k\s}_{\l\m} d\s.  \nn\EEA
The linear part of
equation~(\ref{H4_eq}) has the same form as the linear part of the
corresponding equation obtained for the three-wave
case~(\ref{H3_eq}). Thus, this linear part corresponds to the same
first two terms as in Hamiltonian (\ref{H3_hami}).  Note also
similarity of the quadratic terms in (\ref{H3_eq}) and (\ref{H4_eq}).
Note that (\ref{H4_eq}) correspond to the following Hamiltonian \BEA H
&=& \int A(\k,\l)c_\l c_\k^*d\l d\k+ \frac{1}{2}\int \left[
B(\k,\l)c_{-\l}^*c_\k^*+c.c.\right]d\k d\l \CR &&+ \frac{1}{2}\int
[W^\k_{\l\m}c_\k^*c_\l c_\m +c.c.]d\k d\l d\m+ \nonumber \\
&&+\frac{1}{2}\int T^{\k\l}_{\m\s}c_\k^*c_\l^* c_\m
c_\s\delta^{\k\l}_{\m\s} d\k d\l d\m d\s. \nonumber \\
\label{H4_hamiltonianGEN}
\EEA 
This appears to be a standard Hamiltonian for the wave system with four-wave 
interactions in the presence of spatial inhomogeneity. 
Indeed, the quadratic part (first line)
of this Hamiltonian is the Hamiltonian
(\ref{intr_genHamiltonian}). Cubic term is the three-wave interactions
with the background large scale wave (i.e. four wave interaction where
the role of the fourth wave is assumed by the background wave). Notice
that unlike traditional three wave interactions in a homogeneous
environment, momentum is {\it not} conserved by this term.  This is
the effect of breaking of spatial symmetry by an inhomogeneous
background.  Lastly, the quartic term (third line) is the standard 
four wave interactions Hamiltonian. We show in this paper that the 
quadratic part of this Hamiltonian may be reduced to the novel canonical Hamiltonian for spatially inhomogeneous systems (\ref{canon_inhom}).

In this section we have demonstrated that if the general wave system
is dominated by three-wave or four-wave interactions, and consists of
short scale waves superimposed on known large-scale motion, its
quadratic Hamiltonian is given by the Eq.(\ref{intr_genHamiltonian}).

\section{Preliminaries}
\label{Preliminaries}

In this Section, we set up the stage for formulation of our results.
Here, we give basic definitions, and obtain frequently used formulas.

We  use the following definition of direct and inverse Fourier transforms:
\BEA
\hat{g}(\k)&=&\frac{1}{(2\pi)^d}\int g(\x) e^{-\i \k\cdot\x}d \x,\nonumber\\
g(\x)&=&\int \hat{g}(\k)e^{\i \k\cdot\x}d\k.\nonumber
\EEA
Next, we generalize the Fourier transform to spatially inhomogeneous
systems. In order to do that, we
use a window transform of $g(\x)$:
\BEA
\Gamma[g(\x)]\equiv\tilde{g}(\x,\k) = \frac{1}{(2\pi)^d}\int f(\eps|\x-\x_0|)g(\x_0)e^{-\i \k\cdot\x_0}~d\x_0\label{def_Gabor}.
\EEA
Here, $f(x)$ is an arbitrary fast decaying at infinity window
function.  The parameter $\eps$ is defined by the spatial scales of
the inhomogeneity and the propagating wave-packets in the following manner.  First, we introduce
the characteristic length of inhomogeneity to be of the order
$1/\ep$. Then, we take the width of the window, which is of the order
$1/\eps$, to be much smaller than the characteristic length of
inhomogeneity.  On the other hand, the width of the window is chosen
to be much larger than the wavelength of the waves that propagate in
the inhomogeneous medium, which is of the order $1$.  Therefore, we
have
\BEA
\ep\ll\eps\ll 1.\label{inequal1}
\EEA
The special case when $f(x)=\exp(-x^2)$ is called Gabor
transform~\cite{mallat}.  Note that, when $\eps$ approaches zero,
$f(\eps x)$ approaches the constant function with the value
one. Consequently, the Gabor transform becomes a Fourier transform.
Therefore the Fourier transform can be seen as an averaging over an
infinitely large window.

The inverse of the window transform~(\ref{def_Gabor}) is given by
\BEA
g(\x)=\int\tilde{g}(\x,\k)e^{\i \k\cdot\x}~d\k,\label{def_Gabor_inv}
\EEA
where we have used $f(0)=1$.  We emphasize that
Eq.~(\ref{def_Gabor_inv}) and all the formulas that we obtain below
can be obtained using any fast decaying at infinity window function
and are independent of the particular form of $f(x)$ as long as it is
sufficiently smooth.

Now, we present the formulas for the window transform, which will be
useful later.  First, we express the window transform
$\tilde{g}(\x,\k)$ in terms of the Fourier transform $\hat{g}(\k)$
\BEA
\tilde{g}(\x,\k)=\frac{1}{(\eps)^d}\int
\hf(|\k-\q|/\eps)e^{\i(\q-\k)\cdot\x}\hat{g}(\q)d\q.\label{GaborFourier}
\EEA
Next, we express the Fourier image $\hat{g}(\k)$ in terms of the window
variable $\tilde{g}(\x,\k)$
\BEA
\hat{g}(\k)=\left(\frac{\eps}{\sqrt{\pi}}\right)^d\int\tilde{g}(\x,\k)d\x.\label{FourierGabor}
\EEA
By combining Eqs.~(\ref{GaborFourier}) and~(\ref{FourierGabor}), we obtain
the following formula
\BEA
\tilde{g}(\x,\k)=\left(\frac{1}{\sqrt{\pi}}\right)^d\int
\hf(|\q-\k|/\eps)e^{\i(\q-\k)\cdot\x}\tilde{g}(\x',\q)d\q d\x'.
\label{akxaqxprime}
\EEA
After introducing notations and formulas that will be extensively used below, we proceed to the discussion of the main results of the paper.
\section{The case of nearly-diagonal Hamiltonians.}
\label{Lemma}
\subsection{Formulation and Proof of the Lemma}
Let us start with Hamiltonian (\ref{intr_genHamiltonian}) without off-diagonal
terms, so that $B(\q,\q_1)\equiv 0$. This is a typical
Hamiltonian for linear waves in weakly inhomogeneous media~\cite{ZLF}
expressed in terms of Fourier amplitudes $a_{\q}$ and $ a^*_{\q_1}$ as
\BEA
H=\int\W(\q,\q_1)a_\q a^*_{\q_1}d\q d\q_1,\label{Ham_lemma}
\EEA
with a Hermitian kernel $\W(\q_1,\q)=\W^*(\q,\q_1)$, which is strongly
peaked at $\q-\q_1=0$ ($\W$ has a finite support around $\q\simeq\q_1$).
Therefore, we subsequently assume that there is a small parameter $\ep$
for which
\BEA
\W(\q_1-\q)  =  0, \label{OmegaFinite}
\EEA
when $|\q-\q_1|>\ep$.  A  particular  choice
\BEA
\Omega(\q_1,\q)=\omega(\q_1)\delta(\q-\q_1),
\nn\EEA
leads to the familiar form of the Hamiltonian (\ref{intr_canon}). \\
The equation of motion for $a_\k$ is
\BEA
\i\dot{a}_\k = \frac{\delta H}{\delta a_\k^*} = \int \W_{\q\k}a_\q d\q.
\label{equationofmotion}
\EEA
\begin{lem}\nonumber
Consider the Hamiltonian (\ref{Ham_lemma}) with $\W(\q,\q_1)$ being a
peaked function of $(\q-\q_1)$ (satisfying (\ref{OmegaFinite})) and a
smooth function of $(\q+\q_1)$.  Then there exist a near-canonical
change of variables ${a}_\k\rightarrow\ac_{\k\x}$ such that in the new
variables the equation of motion can be written in the Hamiltonian
form
$$
\i\ddt{\ac_{\k\x}} = \frac{\delta H_f}{\delta \ac^*_{\k\x}},
$$
with the filtered Hamiltonian in a canonical form
\BEA
H_f=\int \ac_{\k\x} [\w_{\k\x}-\x\cdot\nabla_\x\w_{\k\x}+i\{\w_{\k\x},\cdot\}]\ac_{\k\x}^*d\k d\x,\label{HamPois}
\EEA
where $\w_{\k\x}$ is the position dependent frequency related to
$\W(\q,\q_1)$ via the Wigner transform
\BEA
\w_{\k\x}=\int e^{\i \aeb\cdot\x}\W(\k-\aee,\k+\aee)~d\aeb.\label{wkx}
\EEA
\end{lem}
\begin{proof}
In order to obtain the new variables $\ac_{\k\x}$, we first make a
window transform via~(\ref{def_Gabor}), which is then followed by a
near-identical transformation.  The
idea of the proof is to use the peakness of the kernel
$\Omega(\q_1,\q)$.  We make a Taylor expansion around the peak and
then by neglecting the higher order terms we obtain the desired
result.\\

We make a window transform from $a_\k$ to $\at_{\k\x}$ using Eq.~(\ref{GaborFourier}).
Differentiating Eq.~(\ref{GaborFourier}) with respect to time, using the Eq.~(\ref{equationofmotion}), and applying the inverse
formula~(\ref{FourierGabor}) yield
\BEA
\i\ddt{\at_{\k\x}}&=&
\left(\frac{1}{\eps}\right)^d\int \hf(|\k-\q_1|/\eps)e^{\i(\q_1-\k)\cdot\x}\W_{\q\q_1}a_\q d\q d\q_1\nonumber\\
&=&
\left(\frac{1}{\spi}\right)^d\int \hf(|\k-\q_1|/\eps)e^{\i(\q_1-\k)\cdot\x} \W_{\q\q_1}\at_{\q\x_1} {d\q d\q_1 d\x_1}.\label{at_dot1}
\EEA
Let us change variables from $(\q,\q_1)$ to $(\p,\aeb)$ as
\BEA
\q&=&\p-\aee,\label{q}\\
\q_1&=&\p+\aee.\label{q1}
\EEA
Below it will be convenient to use
\BEA
F(\p,\aeb)\equiv\W(\p-\aee,\p+\aee).\label{F}
\EEA
Next, we will approximate the RHS of Eq.~(\ref{at_dot1}) by a
variation of some filtered Hamiltonian $H_f$, i.e., by $\delta
H_f/\delta\at_{\k\x}^*$.  We can rewrite Eq.~(\ref{at_dot1}) as
\BEA
\i\ddt{\at_{\k\x}}=\left(\frac{1}{\spi}\right)^d\int \hf\left(|\k-\p-\aee\right|/\eps) e^{\i(\p+\aee-\k)\cdot\x}F(\p,\aeb) a_{\p-\aee,\x_1}d\p d\aeb d\x_1.\label{at_dot2}
\EEA
Let us make another change of variables $\p\rightarrow \p+\aee$
\BEA
\i\ddt{\at_{\k\x}}=\left(\frac{1}{\spi}\right)^d\int \hf\left(|\k-\p-\aeb\right|/\eps) e^{\i(\p+\aeb-\k)\cdot\x}F(\p+\aee,\aeb) a_{\p,\x_1}d\p d\aeb d\x_1.\label{at_dot3}
\EEA
In order to simplify Eq.~(\ref{at_dot3}), we are going to use the fact
that $\W_{\q\q_1}$ and $\hf(\k)$ are peaked functions of $(\q_1-\q)$
and $\k$, respectively, and fast decaying at infinity.  We also keep
only first order terms in spatial derivatives, neglecting second and
higher order terms. Then, we could write
\BEA
\hf(|\k-\p-\aeb|/\eps)=\hf(|\k-\p|/\eps)+\aeb\cdot\nabla_\p \hf(|\k-\p|/\eps)+\mbox{h.o.t.}\label{ftailor}
\EEA
Similarly, we obtain
\BEA
F(\p+\aee,\aeb)=F(\k+\p-\k+\aee,\aeb)=F(\k,\aeb)+(\p-\k+\aee)\cdot\nabla_\k F(\k,\aeb)+\mbox{h.o.t.}\label{Ftailor}
\EEA
where h.o.t. denotes higher order terms.  Now, we substitute the
expansions~(\ref{ftailor}) and~(\ref{Ftailor}) into
Eq.~(\ref{at_dot2}), and after ignoring higher order terms, we obtain
{
\BEA
\i\ddt{\at_{\k\x}}=&&
\left(\frac{1}{\spi}\right)^d
\CR&&\times
\int\Big(\hf(|\k-\p|/\eps)+\aeb\cdot\nabla_\p 
\hf(|\k-\p|/\eps)\Big)e^{\i(\p+\aeb-\k)\cdot\x}\CR
&&\times\big(F(\k,\aeb)+
(\p-\k+\aee)\cdot\nabla_\k
F(\k,\aeb)\big)\at_{\p,\x_1}d\p d\aeb d\x_1.\nonumber\\
\label{at_dot4}
\EEA
}
Note that, here we have an expansion with two different small
parameters $\ep$ and $\eps$, which obey Eq.~(\ref{inequal1}).

In Appendix~\ref{app1}, we show how to simplify the RHS of
Eq.~(\ref{at_dot4}).  As a result of this simplification, we obtain
\BEA
\i\ddt{\at_{\k\x}}&=&\w_{\k\x}\at_{\k\x}-\x\cdot\nabla_\x\w_{\k\x}\at_{\k\x}+\i\nabla_\x\w_{\k\x}\cdot\nabla_\k\at_{\k\x}-\i\nabla_\k\w_{\k\x}\cdot\nabla_\x\at_{\k\x}
+\underline{\i\nkx\w_{\k\x}(\x\cdot\nabla_{\x}\at_{\k\x})}\nonumber\\
& &+\i/2\nkx\w_{\k\x}\at_{\k\x}+\underline{\nkx\w_{\k\x}\nkx\at_{\k\x}}.\label{combined}
\EEA
Here we have underlined the terms that have two spacial derivatives. 
In the spirit of WKB approximation these terms will later be neglected. 
This equation can be written in the Hamiltonian form
\BEA
\i\ddt{\at_{\k\x}}=\frac{\delta H_f}{\delta\at_{\k\x}^*},
\nn\EEA
where the filtered Hamiltonian takes form
\BEA
H_f&=&\int\Big((\w_{\k\x}-\x\cdot\nabla_\x\w_{\k\x})|\at_{\k\x}|^2+\nkx\w_{\k\x}\at_{\k\x}^*\nkx\at_{\k\x}\CR
& &~~~~+\i\at_{\k\x}^*(\nabla_\x\w_{\k\x}\cdot\nabla_\k\at_{\k\x}-\nabla_\k\w_{\k\x}\cdot\nabla_\x\at_{\k\x}+
\CR &&~~~~~~~
\nkx\w_{\k\x}(\x\cdot\nabla_\x\at_{\k\x})+
1/2\nkx\w_{\k\x}\at_{\k\x})
    \Big)d\k d\x. \label{hHf}
\EEA
Now, we will use the general method of the WKB approximation.  We will
only keep the terms, which are of the first order in a small parameter $\ep$.
In our case, the small parameter $\ep$ characterizes the rate of
spatial change of the position dependent frequency $\w_{\k\x}$ and the
dynamical variable $\at_{\k\x}$.  To apply the WKB approximation to
Eq.~(\ref{combined}), we neglect the terms that have two derivatives
with respect to $\x$ (underlined) because each spatial derivative is of the order $\ep$ small.  As a result, we obtain the
following equation of motion
\BEA
\i\ddt{\at_{\k\x}}=\w_{\k\x}\at_{\k\x}-\x\cdot\nabla_\x\w_{\k\x}\at_{\k\x}+\i\nabla_\x\w_{\k\x}\cdot\nabla_\k\at_{\k\x}-\i\nabla_\k\w_{\k\x}\cdot\nabla_\x\at_{\k\x}
+\i/2 \nkx\w_{\k\x}\at_{\k\x}.\label{combined2}
\EEA
However, Eq.~(\ref{combined2}) becomes non-Hamiltonian. Indeed, the
corresponding functional
\BEA
H_f=\int\Big((\w_{\k\x}-\x\cdot\nabla_\x\w_{\k\x})|\at_{\k\x}|^2+\i\at_{\k\x}^*(\nabla_\x\w_{\k\x}\cdot\nabla_\k\at_{\k\x}-
\nabla_\k\w_{\k\x}\cdot\nabla_\x\at_{\k\x}+\CR
1/2\nkx\w_{\k\x}\at_{\k\x})
    \Big)d\k d\x, \label{hHf2}
\EEA
is not self-adjoint if $\nkx\w_{\k\x}\neq 0$.  Therefore, in order
to obtain canonical equations of motion, another near-canonical change
of variables needs to be performed
\BEA
\at_{\k\x}(t)=s_{\k\x}\ac_{\k\x}(t),\label{new_transform}
\EEA
where $s_{\k\x}$ is some time-independent function to be determined
below.  Note that transformation~(\ref{new_transform}) is canonical if
and only if
\BEA
|s_{\k\x}|^2=1.
\nn\EEA
Therefore, we need to find such $s_{\k\x}$ that the system becomes
Hamiltonian in terms of new variables $\ac_{\k\x}$ and the
transformation~(\ref{new_transform}) is near-canonical, i.e.,
$|s_{\k\x}|\approx 1$.  We substitute Eq.~(\ref{new_transform}) into
Eq.~(\ref{combined2}) to obtain
\BEA
\i s_{\k\x}\ddt\ac_{\k\x}&=&s_{\k\x}\big[(\w_{\k\x}-\x\cdot\nabla_x\w_{\k\x})\ac_{\k\x}+\i\nabla_\x\w_{\k\x}\cdot\nabla_\k\ac_{\k\x}-
\i\nabla_\k\w_{\k\x}\cdot\nabla_\x\ac_{\k\x}\big]\nonumber\\
& &+\ac_{\k\x}[\i\nabla_\x\w_{\k\x}\cdot\nabla_\k s_{\k\x}-\i\nabla_\k\w_{\k\x}\cdot\nabla_\x s_{\k\x}+\i/2 \nkx\w_{\k\x}s_{\k\x}]. \nonumber
\EEA
If we find $s_{\k\x}$ that satisfies the equation
\BEA
\nabla_\k\w_{\k\x}\cdot\nabla_\x s_{\k\x}-\nabla_\x\w_{\k\x}\cdot\nabla_\k s_{\k\x}=\frac{1}{2}s_{\k\x}\nkx\w_{\k\x},\label{eqf}
\EEA
then the equation of motion in the new variables $\ac_{\k\x}$ takes
the canonical form
\BEA
\i \ddt{\ac_{\k\x}}=(\w_{\k\x}-\x\cdot\nabla_\x\w_{\k\x})\ac_{\k\x}-\i\nabla_\k\w_{\k\x}\cdot\nabla_\x\ac_{\k\x}+
\i\nabla_\x\w_{\k\x}\cdot\nabla_\k\ac_{\k\x},\label{ac_dot}
\EEA
with the corresponding Hamiltonian~(\ref{HamPois}).  In order to find
a solution of Eq.~(\ref{eqf}), we make a change of variables
\BEA
g_{\k\x}=2\ln s_{\k\x},\label{gf}
\EEA
to obtain
\BEA
\nabla_\k\w_{\k\x}\cdot\nabla_\x g_{\k\x}-\nabla_\x\w_{\k\x}\cdot\nabla_\k g_{\k\x}=\nkx\w_{\k\x}.\label{eqg}
\EEA
We find the solution of Eq.~(\ref{eqg}) using the method of
characteristics.  The characteristics $(\x(\tau),\k(\tau))$ are given
by the following equations
\BEA
\frac{d\x}{d\tau}&=&\nabla_\k\w_{\k\x},\\
\frac{d\k}{d\tau}&=&-\nabla_\x\w_{\k\x},
\nn\EEA
where $\tau$ is a parameter along the characteristics. Physically
these characteristics correspond to the trajectories (rays) of WKB
wavepackets in the $(\x,\k)$ space.  The solution of Eq.~(\ref{eqg})
is given by
\BEA
g(\x(\tau),\k(\tau))=\int_0^\tau\nkx\w_{\k\x}~d\tau'.
\nn\EEA
Now, we use Eq.~(\ref{gf}) and then Eq.~(\ref{new_transform}) in order
to obtain the new variable $\ac_{\k\x}$ out of the Gabor variable
$\at_{\k\x}$.  The dynamics of the new variable $\ac_{\k\x}$ is
described by the filtered Hamiltonian~(\ref{HamPois}).
\end{proof}

Note that for the special case $\nkx\w_{\k\x}=0$, the Gabor variables
$\at_{\k\x}$ provide a Hamiltonian structure~(\ref{HamPois}).  Then,
in this special case, there is no need in performing the second
transformation~(\ref{new_transform}). We will see below in the
examples that this observation may significantly simplify the
applications of the Lemma.

To describe the statistical properties of spectral energy transfer in the
systems, it is convenient to define a position dependent wave action as
\BEA
\cn_{\k\x}\equiv \langle|\ac_{\k\x}|^2\rangle .\label{nkx}
\EEA
Using this definition, one obtains from (\ref{HamPois}) the familiar
form~(\ref{kinetic}) of the kinetic equation (sometimes called radiative balance
equation) in a weakly inhomogeneous media, by using the  Eq.~(\ref{ac_dot}). 
The resulting equation for the time evolution of $\cn_{\k\x}$ is the 
Eq.~(\ref{kinetic}) of the introduction. 
This is the so-called
waveaction transport which is typical for WKB systems.

\subsection{Relation to the Wigner Transformation}
One can also derive waveaction transport equation (\ref{kinetic}) 
directly from equation of motion (\ref{equationofmotion}), 
without obtaining first the Hamiltonian structure 
(\ref{canon_inhom}). To do this 
we define the Wigner waveaction 
by using the Wigner transformation:
\BEA
\cnw_{\k\x}\equiv\int e^{i\aeb\cdot\x} 
\langle \hat{a}_{\k+\frac{1}{2}\aeb}  \hat{a}_{\k-\frac{1}{2}\aeb}^*
\rangle d\aeb.\nonumber\\
\label{Wigner}
\EEA
Then the Wigner waveaction $\cnw_{\k\x}$ obeys the same kinetic
equation (\ref{kinetic}).  The prove can be found, for example,
in~\cite{VSLvovLectures}.  The idea of the proof is to calculate the
time evolution of waveaction $\cnw_{\k\x}$ by using definition (\ref{Wigner})
and equation of motion (\ref{equationofmotion}). Then one uses
transformation similar to (\ref{q},\ref{q1}), and expands $\cnw_{\k\x}$ using
the smallness of $\aeb$, and finally uses integration by parts to obtain
(\ref{kinetic}) - see \cite{VSLvovLectures} for details.

To address the question on how waveaction $\cnw_{\k\x}$, defined through the
Wigner transformation, is related to the wave action $\cn_{\k\x}$, defined
through the Gabor variables, we substitute (\ref{FourierGabor}) into
the definition of (\ref{Wigner}) to write
$$
\cnw_{\k\x}= 
\Bigg(\frac{\eps}{\sqrt{\pi}}\Bigg)^{2d}
\int \langle \ac_{\k+\m/2,\x'}\ac_{\k-\m/2,x''}^*\rangle
e^{i \m\cdot\x} d\x'd\x'' d\m .
$$ 
Taking into account that $\ac_{\k+\m/2,\x'}$ and $\ac_{\k-\m/2,x''}^*$
are slow functions of $\x'$ and $\x''$, one can neglect this slow
coordinate dependence relative to fast coordinate dependence in the
exponent. This allows to perform $\x'$ and $\x''$ integrations to
obtain a delta-function with respect to the $\m$
argument. Consequently we obtain that these two wave-actions,
(\ref{Wigner}) and (\ref{nkx}) are approximately proportional to each
other:
$$
\cnw_{\k,\x}\propto  \cn_{\k,\x}.
$$
There are several important advantages of our method.  First, it
allows to rigorously write the equation of motion for the field
variable $a_{\k,\x}$ in addition to the transfer equation of
(\ref{kinetic}).  In addition, our approach shows how to derive the
kinetic equation (\ref{kinetic}) to a much broader class of nonlinear
systems, those described by equation (\ref{intr_genHamiltonian}) with
nonzero value of $B(\q,\q_1)\neq 0$, as we show in the next section.
Lastly, Hamiltonian formulation helps to rigorously establish a 
wave turbulence theory and to take into account nonlinear wave
interactions. 

\subsection{Example: Linear Schr\"{o}dinger equation}

Consider a one-dimensional example of a linear Schr\"{o}dinger equation
with a slowly varying potential. This equation is also referred to as 
linearized Gross-Pitaevsky equation. It is used to describe a
formation of the BEC. It is given by
\BEA
\i\dot\psi=-\nabla_{x}^2\psi+U(x)\psi.\label{Schrod}
\EEA
This equation can be written in a Hamiltonian form with the
Hamiltonian given by
\BE
H=\int\left(|\nabla_x\psi|^2+U(x)|\psi|^2\right)~dx.
\EE
In the Fourier space, Eq.~(\ref{Schrod}) becomes
\BE
\i\ddt\psih_k=k^2\psih_k+\int\hat{U}(k-k_1)\psih(k_1)~dk_1,
\EE
with a corresponding Hamiltonian
\BEA
H=\int\Omega(k,k_1)\psih_k\psih_{k_1}^*~dkdk_1,
\nn
\EEA
where $\Omega(k,k_1)=k^2\delta^k_{k_1}+U(k_1-k)$.  Now we can apply
Lemma and find that the position dependent dispersion becomes
\BEA
\w_{kx}=k^2+U(x),
\nn
\EEA
and the corresponding Hamiltonian in terms of Gabor variables takes
the canonical form (\ref{canon_inhom}):
\BEA
H_f=\int \psit_{kx} [\wkx-x\nabx\wkx+i\{\wkx,\cdot\}]\psit_{kx}^*dk dx.
\label{SchrodH}
\EEA
It follows from the Lemma that the Gabor variables provide a canonical
description of the system~(\ref{Schrod}). Indeed, since
$\nabla_k\nabla_x\w_{kx}=0$, we do not need to make a near-identity
transformation~(\ref{new_transform}) in the Lemma. Therefore, it is instructive to obtain the
same result by directly applying Gabor transform to the both sides of
Eq.~(\ref{Schrod}).  We have
\BEA
\Gamma[\nabla_x^2\psi]=\nabla_x^2\psit_{kx}+
2\i k\nabla_x\psit_{kx}-k^2\psit_{kx}\approx2\i k\nabla_x\psit_{kx}
-k^2\psit_{kx}.\label{nabla2}
\EEA
To obtain this equation we have neglected the higher order derivative
of the Gabor variable, since it is  slowly varying in $x$.  Next, we
use the linear expansion of the potential $U(x_0)\approx
U(x)+(x_0-x)\nabla_xU(x)$ to find
\BEA
\Gamma[U(x)\psi]=(U(x)-x\nabla_xU(x))\psit_{kx}+
\i\nabla_xU(x)\nabla_k\psit_{kx}.\label{Upsi}
\EEA
Combining Eq.~(\ref{nabla2}) with Eq.~(\ref{Upsi}), we obtain
\BEA
\i\ddt\psit_{kx}=(k^2+U(x)-x\nabla_xU(x))\psit_{kx}+
\i\nabla_xU(x)\nabla_k\psit_{kx}-2\i k\nabla_x\psit_{kx}.
\nn\EEA
Hamiltonian that corresponds to this equation is given by Eq.~(\ref{SchrodH}).

\subsection{Example: an advection-type system}
Let us consider an advection-type system. For simplicity of
calculations let us restrict our attention to a one dimensional case,
although a general dimensional case can also be considered.  An
advection-type system has a Hamiltonian of the form
\BEA
H=\i\int U(x)\big[\psi(x)\nabla_x\psi^*(x)-\psi^*(x)\nabla_x\psi(x)\big]dx.
\label{ex_Hx}
\EEA
with the corresponding equation of motion
\BEA
\i\dot{\psi}(x)&=&-\i\nabla_x(U(x)\psi(x))-\i U(x)\nabla_x{\psi(x)}
\nonumber\\
&=&-\i(2U(x)\nabla_x\psi(x)+\nabla_x U(x)\psi(x)).\label{ex_eq2}
\EEA
In the Fourier space, this system is described by the Hamiltonian
\BEA
H=\int\Omega(k,k_1)\psih_k\psih_{k_1}^*dkdq_{k_1},\label{ex_Hq}
\nn\EEA
with the kernel
\BEA
\Omega(k,k_1)=(k+k_1)\hat{U}(k_1-k),\label{ex_Omega}
\EEA
After applying Lemma to Hamiltonian~(\ref{ex_Hx}), we obtain the
following canonical form
\BEA
H_f=\int \check{\psi}_{kx} [\wkx-x\nabx\wkx+i\{\wkx,\cdot\}]\check{\psi}_{kx}^*dk dx,\label{SchrodH_ad}
\EEA
where $\check{\psi}$ are the new variables.
\BEA
\w_{kx}=2kU(x),\label{ex_wkx}
\EEA
is a position dependent frequency.  Note that in this case we have
$\nabla_k\nabla_x\w_{kx}\neq 0$ and the near-canonical change of
variables given by Eq.~(\ref{new_transform}) had to be performed.

We can also obtain the same result by directly applying the Gabor
transform to Eq.~(\ref{ex_eq2}).
Using the slow dependence of $U(x)$ on $x$ (disregarding the second
derivative and higher), we obtain
\BEA
\Gamma[U(x)\nabla_x\psi(x)]\approx
\CR
\big(U(x)-x\nabla_xU(x)+\i\nabla_xU(x)\nabla_k\big)(\nabla_x+\i k)\psit_{kx}.\label{ex_tmp1}
\EEA
Similarly, we have
\BEA
\Gamma[\nabla_xU(x)\psi(x)]\approx\nabla_xU(x)\psit_{kx}.\label{ex_tmp2}
\EEA
Substituting Eqs.~(\ref{ex_tmp1}) and~(\ref{ex_tmp2}) into
Eq.~(\ref{ex_eq2}), we obtain
\BEA \i\ddt{\psit}_{kx}=\big(-\i 2U(x)\nabla_x+2kU(x)+2\i x\nabla_x
U(x)\nabla_x-2xk\nabla_xU(x)+2\nabla_xU(x)\nabla_{kx}+
\CR
2\i\nabla_xU(x)+2\i\nabla_xU(x)k\nabla_k -\i\nabla_x
U(x)\big)\psit_{kx}.\nonumber\\ \label{ex_eq3} \EEA
Using Eq.~(\ref{ex_wkx}), we rewrite Eq.~(\ref{ex_eq3}) as
\BEA
\i\ddt{\psit}_{kx}=\left(\w_{kx}-x\nabla_x\w_{kx}+\underline{\nabla_k\nabla_x\w_{kx}\nabla_k\nabla_x}+\i(\nabla_x\w_{kx}\nabla_k-
\nabla_k\w_{kx}\nabla_x)+
\right. \CR \left.
\underline{\i x\nabla_k\nabla_x\w_{kx}\nabla_x}+
\frac{1}{2}\i\nabla_k\nabla_x\w\right)\psit_{kx}.\label{ad_psit}
\EEA
As in the Lemma, we neglect the higher order terms with the two
derivatives over $x$ (underlined in Eq.~(\ref{ad_psit})).  In order
to obtain the canonical form of the equation of motion, we need to
make a near-canonical transformation
\BEA
\psih_{kx}=f \psic_{kx},
\nn\EEA
where $f$ satisfies
\BEA
\nabla_k\w_{kx}\nabla_xf-\nabla_x\w_{kx}\nabla_kf=
\frac{1}{2}f\nabla_k\nabla_x\w_{kx}.\label{fw}
\EEA
For this special case, we obtain
\BEA
U(x)\nabla_xf=\nabla_xU(x)\left(\frac{1}{2}f+k\nabla_kf\right).\label{f}
\EEA
We have to find the solution for Eq.~(\ref{f}) such that $f\rightarrow
1$ when $\nabla_xU(x)\rightarrow 0$.  Therefore, we need to find a
solution in the form $f=1+g$, where $g$ satisfies $|g(k,x)|\ll 1$.
Let us try to find a solution in the form $f=f(x)$, i.e., independent
of $k$.  Then we have
\BEA
f(x)=C\sqrt{U(x)},
\nn\EEA
where $C$ is an arbitrary constant.  We expand $U(x)$ around some
point of reference $x_0$ as
\BEA
U(x)\approx U(x_0)+(x-x_0)\nabla_xU(x_0).
\nn\EEA
Let us choose the constant to be $C=1/\sqrt{U(x_0)}$ then we obtain
\BE
f\approx 1+\frac{1}{2}(x-x_0)\frac{\nabla_xU(x_0)}{U(x_0)}.
\nn\EE
If $\nabla_xU(x_0)\sim\ep$ and $|x-x_0|\ll 1/\ep$ then $f\approx 1$
and the transformation is near-canonical.

\section{General case of waves in weakly-inhomogeneous media}
\label{Theorem}
\subsection{Theorem}
The result of the Lemma can be generalized onto a much broader class
of Hamiltonian given by (\ref{intr_genHamiltonian}).
Thus let us consider (\ref{intr_genHamiltonian}) with both $A$ and $B$ being peaked functions of $\q-\q_1$,
which corresponds to waves on a weakly inhomogeneous background.  Note that
\BEA A(\q_1,\q)=A^*(\q,\q_1),
\label{cond_A}
\EEA
 because the Hamiltonian is Hermitian.  Moreover
\BEA
B(-\q,-\q_1)=B(\q_1,\q).\label{cond_B}
\EEA
Condition~(\ref{cond_B}) does not really restrict our choice of the
coefficient $B(\q,\q_1)$.  Indeed, we can consider any function $B$
and then represent it as a sum of two components
\BEA
B(\q,\q_1)=B^{'}(\q,\q_1)+B^{''}(\q,\q_1),\label{expansionB}
\EEA
where
\BEA
B^{'}(\q,\q_1)=\frac{1}{2}(B(\q,\q_1)+B(-\q_1,-\q)),\nonumber\\
B^{''}(\q,\q_1)=\frac{1}{2}(B(\q,\q_1)-B(-\q_1,-\q)).\nonumber
\EEA
When we plug Eq.~(\ref{expansionB}) into
Hamiltonian~(\ref{intr_genHamiltonian}), the part of the integral with
$B^{''}$ vanishes.

From now on, we will omit indices wherever it does not lead to  confusion
and denote $a\equiv \ac_{\k,\x}$ and $a_{-}\equiv \ac_{-\k,\x}$.
Also, we introduce some convenient notations, which we will use
throughout the rest of the paper.  For any function $\varphi(\k)$, we
denote its even part as $\varphi_{ev}$ and its odd part $\varphi_{od}$
\BEA \varphi_{ev}&=&\frac{1}{2}(\varphi+\varphi_{-}),\nonumber\\
\varphi_{od}&=&\frac{1}{2}(\varphi-\varphi_{-}).\nonumber \EEA
We now are ready to formulate the main theorem of this paper:
\begin{thm}
Consider a Hamiltonian (\ref{intr_genHamiltonian}) with $A(\q,\q_1)$ and
$B(\q,\q_1)$ being peaked functions of $(\q-\q_1)$ with the same
parameter $\ep$, i.e., $A(\q,\q_1)=0$ and $B(\q,\q_1)=0$ when
$|\q-\q_1|>\ep$.  Suppose that conditions~(\ref{cond_A})
and~(\ref{cond_B}) are satisfied.  Let us introduce new notations
\BEA
\mu&\equiv&\int A(\k-\aee,\k+\aee) e^{\i\aeb\cdot\x} d\aeb, \CR
\lambda&\equiv&\int B(\k-\aee,\k+\aee) 
e^{\i\aeb\cdot\x}d\aeb\label{lambda_def},\CR
\nu&\equiv&\Re[\lambda]\label{nu_def},\CR
\widetilde{\nu}&\equiv&\Im[\lambda].
\nn\EEA
Suppose that Hamiltonian~(\ref{intr_genHamiltonian}) has a dominant
diagonal part, i.e. $\mu_{ev}>\nu$ and $\tilde{\nu}=O(\varepsilon)$.
Then, there exists a new-canonical change of variables from $a_\k$ to
$c_{\k\x}$ and the evolution of system~(\ref{intr_genHamiltonian}) can be
approximately described by the following equation of motion
\BEA
\i\ddt{c_{\k\x}} = \frac{\delta H_f}{\delta c_{\k\x}^*},
\nn\EEA
with a filtered Hamiltonian
\BEA
H_f=\int c_{\k\x}[\w-\x\cdot\nabla_\x\w+\i\{\w,\cdot\}]
                        c_{\k\x}^*d\k d\x.\label{finalHam}
\EEA
The position dependent frequency is given by the formula
\BEA
\omega=\mu_{od}+\sqrt{\mu_{ev}^2-\nu^2}.
\nn\EEA
The transformation from the Fourier variables $a_\k$ to new variables $c_{\k\x}$ is given in the proof.
\end{thm}
\begin{proof}
The proof consists of three main steps. In order to diagonalize
Hamiltonian~(\ref{intr_genHamiltonian}), we
\begin{enumerate}
\item apply the Lemma to simplify the Hamiltonian using the peakness
of the kernels,
\item perform Bogolyubov transformation to diagonalize the $O(1)$ part
of the Hamiltonian,
\item make a near-identity canonical transformation to diagonalize the
$O(\varepsilon)$ part of the Hamiltonian.
\end{enumerate}
\textbf{Step 1:} Applying Lemma.
\\
Similarly to Eq.~(\ref{HamPois}), we can write a filtered Hamiltonian
for Eq.~(\ref{intr_genHamiltonian}) as
\BEA
H_f^{(1)}&=&\int \ac(\mu-\x\cdot\nabla_\x\mu+\i\{\mu,\cdot\})\ac^*d\k d\x+
\nonumber \\ &&
\frac{1}{2}\int [\ac [\lambda-\x\cdot\nabla_{\x}\lambda+\i\{\lambda,\cdot\}]\ac_{-}d\k d\x+c.c.],\nonumber \\
\label{genHamPois}
\EEA
here, as usual c.c. stands for complex conjugate.  From
property~(\ref{cond_B}) and definition~(\ref{lambda_def}) it follows
that $\lambda(-\k,\x)=\lambda(\k,\x)$.
\\
\textbf{Step 2:} Bogolyubov transformation.
\\ In this step we apply the usual Bogolyubov transformation. 
Before
doing that notice that Hamiltonian~(\ref{genHamPois}) consists of two
parts
\BEA
H_f^{(1)}=H_{f,1}^{(1)}+H_{f,\ep}^{(1)},\nonumber
\EEA
where
\BEA
H_{f,1}^{(1)}&=&\int\mu |\ac|^2d\k d\x+\frac{1}{2}
\int \nu [\ac\ac_-+\ac^*\ac_-^*]d\k d\x,\nonumber \\
H_{f,\ep}^{(1)}&=&H_f^{(1)}-H_{f,1}^{(1)},\nonumber \\
\label{H1e}
\EEA
are correspondingly $O(1)$ and $O(\ep)$ parts of $H_f^{(1)}$.  In Step
2, we diagonalize the $O(1)$ part using the following linear
transformation
\BEA
\ac=u_{\k\x}b + v_{\k\x}b^*_-.\label{ZLF_transform}
\EEA
It was shown in ~\cite{ZLF} that transformation~(\ref{ZLF_transform})
is canonical if the following conditions are satisfied:
\BEA
|u_{\k\x}|^2 - |v_{\k\x}|^2 = 1,\nonumber \\
u_{\k\x} v_{-\k,\x} = u_{-\k,\x} v_{\k\x}.\nonumber
\EEA
Let us follow~\cite{ZLF} and choose
\BEA
u_{\k\x}=\cosh(\xi_{\k\x}),\label{u_def} \\
v_{\k\x}=\sinh(\xi_{\k\x}),\label{v_def}
\EEA
where $\xi_{\k\x}$ is real and even, but otherwise arbitrary function.
Then under change of variables given by Eq.~(\ref{ZLF_transform}),
$H_{f,1}^{(1)}$ becomes
\BEA
H_{f,1}^{(1)}
&=&\int \Big[\mu\cosh^2(\xi)+\mu_-\sinh^2(\xi)+2\nu\sinh(\xi)\cosh(\xi)\Big]
|b|^2d\k d\x\nonumber \\
&+&\frac{1}{2}\int\Big[(\mu+\mu_-)\sinh(\xi)\cosh(\xi)+\nu\big(\cosh^2(\xi)+\sinh^2(\xi)\big)\Big](bb_-+b^*b_-^*)d\k d\x.\nonumber
\EEA
Denote the expression in square brackets multiplying  $|b|^2$ as $\w$:
$$\omega =  \mu\cosh^2(\xi)+\mu_-\sinh^2(\xi)+
2\nu\sinh(\xi)\cosh(\xi).
$$
Using trigonometric formulas for hyperbolic functions, we obtain
\BEA \w=\muev\cosh(2\xi)+\muod+\nu\sinh(2\xi).\label{omega3} \EEA
In order to diagonalize $H_{f,1}^{(1)}$, we require that the
following condition in satisfied
\BEA
\muev\sinh(\xi)\cosh(\xi)+\nu\Big(\cosh^2(\xi)+\sinh^2(\xi)\Big)=0.
\label{O1diag}
\EEA
This condition is equivalent  to
\BEA \tanh(2\xi)=-\frac{\nu}{\muev}.\label{th}
\EEA
Since $\mu_{ev}>\nu$, we can choose $\cosh(2\xi)$ to be positive and,
therefore, we have
\BEA
\cosh(2\xi)&=&\frac{\mu_{ev}}{\sqrt{\mu_{ev}^2-\nu^2}},\label{coshu}\\
\sinh(2\xi)&=&-\frac{\nu}{\sqrt{\mu_{ev}^2-\nu^2}}.\label{sinhv}
\EEA
In Appendix~\ref{app_xi}, we find the expression for $\xi$.
Resolving Eq.~(\ref{omega3}) together with Eqs.~(\ref{coshu}) and~(\ref{sinhv}), we obtain
\BEA
\omega&=&\wev + \wod,\nonumber \\
\wod&=&\muod,\nonumber \\
\wev&=&\sqrt {\muev^2-\nu^2}.\nonumber
\label{wpm}
\EEA
Therefore, we have diagonalized $H_{f,1}^{(1)}$ to the form
\BEA
H_{f,1}^{(1)}=\int b\w b^*d\k d\x.\label{H1afterBogolyubov}
\EEA
Next, we consider the $O(\varepsilon)$ part of the filtered
Hamiltonian.  In Appendix~\ref{app_Bogolyubov}, we show that
Bogolyubov transformation~(\ref{ZLF_transform}) transforms
$H_{f,\ep}^{(1)}$ to the form
\BEA
H_{f,\ep}^{(1)}&=&\int b\big(-\x\cdot\nabla_{\x}\w+i\{\w,\cdot\}\big)b^*+
\left(\sigma bb_-+\frac{\i\muev^2}{2\nu}b\left\{\varphi,b_-\right\}+c.c.\right)d\k d\x,\nonumber \\
\label{HeafterZLF}
\EEA
where
\BEA
\sigma
&=&\frac{\muev^2}{2\nu}\left(\x\cdot\nabla_{\x}\frac{\omega_{ev}}{\muev}\right)+\frac{i}{2}\{\muod,\xi\}+\frac{i}{2}\widetilde{\nu}\nonumber\\
\varphi&=&\sqrt{1-\frac{\nu^2}{\muev^2}}=\frac{\omega_{ev}}{\mu_{ev}}.\nonumber
\EEA
Combining Eqs.~(\ref{H1afterBogolyubov}) and~(\ref{HeafterZLF}), we
finally obtain Hamiltonian in the form
\BEA
H_f^{\ep}&=&\int b\big(\w-\x\cdot\nabla_{\x}\omega+i\{\omega,\cdot\}\big)b^*+
\left(\sigma bb_-+\frac{\i\muev^2}{2\nu}b\left\{\varphi,b_-\right\}+c.c.\right)d\k d\x.\label{HafterZLF}
\EEA
\textbf{Step 3:} Near-identity transformation.
\\
In Step 2, we diagonalized $O(1)$ part but not all of the
$O(\varepsilon)$ part.  In order to diagonalize complete Hamiltonian,
we use the near-identity transformation.  This near-identity
transformation changes variables from $b_{\k\x}$ to $c_{\k\x}$ by the
following rule
\BEA
b_{\k\x}=c_{\k\x}+\a_\k c_{-\k,\x}^*+\b_\k\{\g_\k,c_{-\k,\x}^*\},\label{nit}
\EEA
where, we assume that $\bk$ and $\gk$ are $O(1)$ terms and $\alk$ and
$\big(\bk\{\gk,c_{-\k}^*\}\big)$ are $O(\varepsilon)$ which makes our
transformation indeed near identical.  Note that $\alk$, $\bk$, and
$\gk$ are functions of both $\k$ and $\x$. Nevertheless, for
simplicity of notation, we do omit the dependence on $\x$, since it
would only unnecessarily pollute the notations.  In
Appendix~\ref{app_canonicity}, we derive the canonicity conditions for
transformation~(\ref{nit}).  In turns out that
transformation~(\ref{nit}) is canonical if the following conditions
are met
\BEA
\b_\k&=&\bmk,\nonumber\\
\g_\k&=&\gmk,\nonumber\\
\alod&=&\frac{1}{2}\{\gk,\bk\}.\nonumber\\
\label{cancon}
\EEA
Among the coefficients $\ak$, $\bk$ and $\gk$ that satisfy the
canonicity conditions we have to choose those that will diagonalize
the $O(\varepsilon)$ part.  In Appendix~\ref{app_nit}, we show that
such coefficients become
\BEA
\ak&=&-\frac{\sigma_\k^*}{\wev}-\frac{\bk}{2\wev}\left\{\wod,\frac{\w_{ev}}{\mu_{ev}}\right\},\nonumber\\
\bk&=&\frac{i\muev^2}{2\nu\wev},\nonumber\\
\gk&=&\frac{\omega_{ev}}{\mu_{ev}}.\nonumber\\
\label{diagcon3}
\EEA
Note that these conditions are
in full correspondence with the canonicity conditions~(\ref{cancon}).
The Hamiltonian in new variables up to $O(\varepsilon)$ order is
\BEA
H_f=\int c[\w-\x\cdot\nabla_\x\w+i\{\w,\cdot\}]c^*d\k d\x.\nonumber
\EEA

This completes the proof of the main result of this paper.
\end{proof}

\subsection{Example: Nonlinear Schr\"{o}dinger Equation --- waves on condensate.}
Bose-Einstein Condensate (BEC) is a state of matter that arises in
dilute gases with large number of particles at very low
temperatures~\cite{Bose,Einst1,Einst2,Anderson,Bradley,Davis}.  BEC
can be described by the Nonlinear Schr\"{o}dinger equation (also known as
Gross-Pitaevskii equation~\cite{Pitaebsky}).  Here, we apply the
theorem to this well studied model.  The evolution of the state
function $\psi$ is described by the following equation
\BEA
\i\frac{\partial \psi}{\partial t}+\triangle\psi-|\psi|^2\psi+\kappa(t)\psi=0,\label{nls_psi_x}
\nn\EEA
with a corresponding Hamiltonian
\BEA
H=\int(|\nabla\psi|^2+\frac{1}{2}|\psi|^4-\kappa(t)|\psi|^2)d\x.\label{nls_H_x}
\nn\EEA
The term $\kappa(t)\psi$ is introduced for convenience as will become
clear later.  Following~\cite{BEC}, let us consider the
amplitude-phase representation of the order parameter
$\psi$:
\BEA
\psi=Ae^{i\varphi}\label{nls_psi_Ap}.
\nn\EEA
Now, we introduce Hamiltonian momentum
\BEA
p=2A\varphi,\label{nls_p_def}
\EEA
and rewrite Eq.~(\ref{nls_psi_x}) in terms of new canonical variables $A$ and
$p$ as
\BEA
A_t&=&\frac{\delta H}{\delta p},\\
p_t&=&-\frac{\delta H}{\delta A},
\nn\EEA
where
\BEA
H=\int\left((\nabla A)^2+\frac{1}{2}A^4-\kappa(t)A^2+\frac{1}{4}\left(\nabla p-\frac{p\nabla A}{A}\right)^2\right)d\x.\label{nls_H_Ap}
\EEA
Let us consider weak perturbations on background of a strong
condensate,
\BEA
A=A^{(0)}+A^{(1)},\ \ p=p^{(0)}+p^{(1)},\ \ |A^{(1)}|\ll |A^{(0)}|.\label{nls_pert}
\EEA
We now choose $\kappa(t)=(A^{(0)})^2$ which gives us
$p^{(0)}\sim\varepsilon$.  Substituting Eq.~(\ref{nls_pert}) into
Eq.~(\ref{nls_H_Ap}) we have
\BEA
H=H_0+H_2+H_3,
\nn\EEA
where the subscripts denote the order of the term with respect to
perturbation amplitudes.  Since in this paper we study the linear
dynamics, we only consider the quadratic part of the Hamiltonian
\BEA
H_2=\int\left(\left(\nabla A^{(1)}\right)^2+(A^{(0)})^2(A^{(1)})^2+
\frac{1}{4}(\nabla p)^2+\frac{1}{2}\frac{p^{(0)}}{A^{(0)}}\nabla p\cdot\nabla A^{(1)}+\frac{1}{4}p\left(\nabla\ln A^{(0)}\right)\cdot\nabla p\right)d\x.
\nn\EEA
Here, we used the fact that the spatial derivative adds one order in
$\varepsilon$ and we neglected the terms of the order two and higher.

In order to apply the theorem to $H_2$ we first transform to Fourier
space and then switch to normal variables.  Let us denote
$R=\left(A^{(0)}\right)^2$, $S=p^{(0)}/A^{(0)}$ and $T=\nabla_x\ln
A^{(0)}$. We have $R=O(1)$ and $S,T=O(\varepsilon)$.  Transforming
$H_2$ into Fourier space we obtain
\BEA
H_2=\int\left((\k_1\cdot\k_2\delta_2^1+R_{2-1})A_1A^*_2+\frac{1}{2}S_{2-1}\k_1\cdot\k_2p_1A^*_2+\frac{1}{4}
\Big(\k_1\cdot\k_2\delta_2^1-T_{2-1}\k_1\cdot(\k_2-\k_1)\Big)p_1p^*_2\right)d12,
\nn\EEA
where we used the following simplified notations: $1\equiv \k_1$,
$2\equiv \k_2$ and subscript $2-1\equiv \k_2-\k_1$.  Next, we switch
to normal variables using the transformation
\BEA
A_\k&=&\frac{1}{\sqrt{2}}(a_\k+a^*_{-\k}),\\
p_\k&=&-\frac{i}{\sqrt{2}}(a_\k-a^*_{-\k}).
\nn\EEA
In normal variables, $H_2$ reads
\BEA
H_2&=&\int\Bigg(\left(\frac{5}{4}\k_1\cdot\k_2\delta_2^1+R_{2-1}+\frac{1}{8}T_{2-1}(\k_2-\k_1)^2\right)a_1a^*_2+\nonumber\\
&+&\frac{1}{2}\left(\frac{3}{4}\k_1\cdot\k_2\delta_2^1+R_{2-1}-\frac{\i}{2}S_{2-1}\k_1\cdot\k_2+\frac{1}{4}T_{2-1}\k_1\cdot(\k_2-\k_1)\right)a_1a_{-2}+c.c.\Bigg)d12.
\nn\EEA
The only part of the coefficient in the second parenthesis we are
interested in is the one that satisfies Eq.~(\ref{cond_B}):
\BEA
\frac{3}{4}\k_1\cdot\k_2\delta_2^1+R_{2-1}-\frac{\i}{2}S_{2-1}\k_1\cdot\k_2+\frac{1}{8}T_{2-1}(\k_2-\k_1)^2.
\nn\EEA
Since $A^{(0)}$ and $p^{(0)}$ are slowly varying functions of $\x$, so
are $R(\x)$, $S(\x)$ and $T(\x)$.  Therefore, their Fourier transforms
are peaked around zero making the terms proportional to
$T_{2-1}(\k_2-\k_1)^2$ of the second order in $\varepsilon$, which can
be neglected.  Finally, we can write down the Hamiltonian in the form
given in Eq.~(\ref{intr_genHamiltonian})
\BEA
H_2=\int \left(A(\k_1,\k_2)a_1a^*_2+\frac{1}{2}(B(\k_1,\k_2)a_1a^*_2+c.c.)\right)d12,\label{nls_genHamiltonian}
\EEA
where
\BEA
A(\k_1,\k_2)&=&\frac{5}{4}\k_1\cdot\k_2\delta_2^1+R_{2-1}\\
B(\k_1,\k_2)&=&\frac{3}{4}\k_1\cdot\k_2\delta_2^1+R_{2-1}-\frac{\i}{2}S_{2-1}\k_1\cdot\k_2.
\nn\EEA
In terms of window transformations, which we denote here as $a$ the
Hamiltonian reads
\BEA
H_f=\int a (\mu-\x\cdot\nabla\mu+\i\{\mu,\cdot\})a^*d\k d\x+\frac{1}{2}\int [a [\lambda-\x\cdot\nabla\lambda+i\{\lambda,\cdot\}]a_{-}d\k d\x+c.c.],\nonumber
\EEA
where
\BEA
\mu&=&\int e^{\i\aeb\cdot\x}A(\k-\aee,\k+\aee)d\aeb=\frac{5}{4}\k^2+R(\x),\\
\lambda&=&\frac{1}{2}\int e^{\i\aeb\cdot\x}B(\k-\aee,\k+\aee)d\aeb=\frac{3}{4}\k^2+R(\x)-\frac{\i}{2}\k^2S(\x)+\frac{1}{2}\k\cdot\nabla S(\x).
\nn\EEA
Up to the first order in $\varepsilon$, we have
\BEA
\nu&=&\frac{3}{4}\k^2+R(\x),\\
\tilde{\nu}&=&-\frac{1}{2}\k^2S(\x).
\nn\EEA
Here, $\mu$ is an even function of $\k$ which means that $\muev=\mu$
and $\muod=0$.  Then, the position dependent frequency of the small
perturbations in the presence of the condensate becomes
\BEA
\omega=\sqrt{\mu^2-\nu^2}=|\k|\sqrt{R(\x)+\k^2}. \nonumber
\EEA
Bogolyubov's transformation, $a=ub+vb^*_-$, is given by the following
coefficients
\BEA
u&=&\frac{\mu}{\sqrt{\mu^2-\nu^2}}=\frac{5\k^2+4R(\x)}{4|\k|\sqrt{\k^2+R(\x)}},\nonumber\\
v&=&-\frac{\nu}{\sqrt{\mu^2-\nu^2}}=-\frac{3\k^2+4R(\x)}{4|\k|\sqrt{\k^2+R(\x)}}.\nonumber
\EEA
In terms of variables $b$ the Hamiltonian takes the following form
\BEA
H_f&=&\int b(\omega-\x\cdot\nabla_{\x}\omega+\i\{\omega,\cdot\})b^*d\k d\x+
\left(\int\sigma bb_-d\k d\x+\frac{\i}{2}\int\left[\frac{b\muev^2}{\nu}\left\{\varphi,b_-\right\}\right]d\k d\x+c.c.\right),\label{fpu_step2}
\EEA
where
\BEA
\sigma&=&\frac{\mu^2}{2\nu}\x\cdot\nabla\sqrt{1-\frac{\nu^2}{\mu^2}}+\frac{\i}{2}\tilde{\nu}=-\frac{\k^2}{4}\left(\frac{\x\cdot\nabla R(x)}{|\k|\sqrt{\k^2+R(\x)}}+
\i S(\x)\right),\nonumber\\
\varphi&=&\sqrt{1-\frac{\nu^2}{\mu^2}}=\frac{4|\k|\sqrt{\k^2+R(\x)}}{5\k^2+4R(\x)}.\nonumber
\EEA
Finally, we perform the near-identity transformation $b=c+\alpha
c^*_-+\beta\{\gamma,c^*_-\}$ where
\BEA \alpha&=&\frac{\x\cdot\nabla
R(\x)}{4(\k^2+R(\x))}-\frac{\i|\k|S(\x)}{4\sqrt{\k^2+R(\x)}},\nonumber\\
\beta&=&\frac{\i(5\k^2+4R(\x))^2}{8|\k|\sqrt{\k^2+R(\x)}(3\k^2+4R(\x))},\nonumber\\
\gamma&=&\frac{4|\k|\sqrt{\k^2+R(\x)}}{5\k^2+4R(\x)}.\nonumber \EEA
The resulting Hamiltonian attains the canonical form (\ref{canon_inhom}).

\section{Conclusions}
We have studied the dynamical behavior of the linearized spatially
inhomogeneous Hamiltonian wave systems.  The canonical transformation
from the Fourier variables to the new spatially-dependent variables is
found for the general class of the quadratic Hamiltonians.  In the new
variables, the linearized dynamics is governed by the canonical
diagonal Hamiltonian with the spatially-dependent dispersion relation.
The waveaction transport equation which coresponds to this Hamiltonian
has form (\ref{kinetic}) which is typical to WKB formalism. It was
previously obtained for some specific examples, e.g. in plasmas
\cite{rudak} and geophysical waves \cite{dnz,janssen}.  In this paper,
we have given several representative examples illustrating the general
results, such as the Nonlinear Schr\"{o}dinger Equation without and
with condensate and an advective-type system. Further possible areas of
application of this formalism include water waves on lakes with
variable depth or/and presence of variable mean flow, internal waves
in media with variable stratification, plasma waves on profiles with
variable density, geophysical waves in media with variable background
rotation rates, etc.

The new Hamiltonian formalism that is presented in this paper should
be crucial for extending the WT theory to the spatially inhomogeneous
systems.  In the spatially homogeneous systems, quadratic term in the
Hamiltonian corresponds to the first term in Eq.~(\ref{finalHam}).
Effect of space inhomogeneity leads to the appearance of the
derivative terms in the Hamiltonian, which correspond to the slow
dynamics along the rays in the $(\bf{k,x})$-space.  This effect will
lead to an interesting interplay of inhomogeneity and non-linearity in
wave turbulence systems. More specifically, linear dispersion relation
becomes spatially dependent. Consequently, the resonance conditions
change as waves propagate through inhomogeneous environment.  As a
result of this, waves will remain in resonance for a \textit{limited}
amount of time, or the members of the resonant triads will change from
position to position. The physical implication of this effect may be
the weakened flux of energy or other conserved quantities through the
wavenumber space. Other effect may be an effective broadening of the
resonances, as resonances will be altered from place to place, so the
wavepacket propagating through the inhomogeneous environment will be
affected by averaged dispersion relation.  Another potentially
interesting effect is an effective three-wave interactions in a
four-wave weak turbulence systems, where the role of fourth wave is
played by inhomogeneity.

In order to develop a Wave Turbulence theory for spatially
inhomogeneous systems, the kinetic equation has to be obtained for the
cases with such finite-time wave resonances.  This is an exciting task
for the future work.
\paragraph*{Acknowledgments}
Y.~L. was supported by NSF CAREER grant, DMS 0134955.  

\appendix
\section{Window transform}
\label{app1}
Let us consider the RHS of Eq.~(\ref{at_dot3}) term by term.\\
1) $\bf{f_0F_0}$\\
\BEA
& &\left(\frac{1}{\spi}\right)^d\int \hf((\k-\p)/\eps)e^{\i(\p+\aeb-\k)\cdot\x}F(\k,\aeb)\at_{\p,\x_1}d\p d\aeb d\x_1\nonumber\\
& &=\left(\frac{1}{\spi}\right)^d\int \hf((\k-\p)/\eps)e^{\i(\p-\k)\cdot\x}\at_{\p,\x_1}d\p d\x_1\int F(\k,\aeb)e^{\i\aeb\cdot\x}d\aeb=\w_{\k\x}\at_{\k\x},\label{f0F0}
\EEA
2) $\bf{f_1F_0}$\\
\BEA
& &\left(\frac{1}{\spi}\right)^d\int\aeb\cdot\nabla_\p \hf((\k-\p)/\eps)e^{\i(\p+\ae-\k)\cdot\x}F(\k,\ae)\at_{\p,\x_1}dpd\aeb d\x_1\nonumber\\
& &=-\left(\frac{1}{\spi}\right)^d\int \hf((\k-\p)/\eps)\i \aeb\cdot\x e^{\i(\p+\aeb-\k)\cdot\x} F(\k,\aeb)\at_{\p,\x_1} d\p d\aeb d\x_1\nonumber\\
& &~~-\left(\frac{1}{\spi}\right)^d\int \hf((\k-\p)/\eps)e^{\i(\p+\aeb-\k)\cdot\x} F(\k,\aeb)\aeb\cdot\nabla_\p\at_{\p,\x_1}d\p d\x_1d\aeb\nonumber\\
& &=-\x\cdot\nabla_\x\w_{\k\x}\at_{\k\x}+\i\nabla_\x\w_{\k\x}\cdot\nabla_\k\at_{\k\x},\label{f1F0}
\EEA
3) $\bf{f_0F_1}$\\
\BEA
& &\left(\frac{1}{\spi}\right)^d\int \hf((\k-\p)/\eps)e^{\i(\p+\aeb-\k)\cdot\x}(\p-\k+\aee)\cdot\nabla_\k F(\k,\aeb)\at_{\p,\x_1} d\p d\aeb d\x_1\nonumber\\
& &=-\i\left(\frac{1}{\spi}\right)^d\int \hf((\k-\p)/\eps)e^{\i(\p-\k)\cdot\x}\i(\p-\k)\at_{\p,\x_1}d\p d\x_1\cdot\int e^{\i\aeb\cdot\x}\nabla_\k F(\k,\aeb)d\aeb\nonumber\\
& &+\left(\frac{1}{\spi}\right)^d\int \hf((\k-\p)/\eps)e^{\i(\p-\k)\cdot\x}\at_{\p,\x_1}d\p d\x_1(-\i)\int \i/2 e^{\i\aeb\x}\aeb\cdot\nabla_\k F(\k,\aeb)d\aeb\nonumber\\
& &=-\i\nabla_\k\w_{\k\x}\cdot\nabla_\x\at_{\k\x}-\i/2(\nabla_{\k}\cdot\nabla_{\x})\w_{\k\x}\at_{\k\x},\label{f0F1}
\EEA
4) $\bf{f_1F_1}$\\
\BEA
& &\int\aeb\cdot\nabla_\p \hf((\k-\p)/\eps)e^{\i(\p+\aeb-\k)\cdot\x}(\p-\k)\cdot\nabla_\k F(\k,\aeb)\at_{\p,\x_1}d\p d\aeb d\x_1\nonumber\\
& &=\int \hf((\k-\p)/\eps)\i e^{\i(\p-\k)\cdot\x}\at_{\p,\x_1}\x\cdot(\p-\k)d\p d\x_1\int-e^{\i\aeb\cdot\x}\aeb\cdot\nabla_\k F(\k,\aeb)d\aeb\nonumber\\
& &+\int \hf((\k-\p)/\eps)e^{\i(\p-\k)\cdot\x}\at_{\p,\x_1}d\p d\x_1\int-e^{\i \aeb\cdot\x}\aeb\cdot\nabla_\k F(\k,\aeb)d\aeb\nonumber\\
& &+\int \hf((\k-\p)/\eps)e^{\i(\p-\k)\cdot\x}\nabla_\p\at_{\p,\x_1}\cdot(\p-\k)d\p d\x_1\int-e^{\i\aeb\cdot\x}\aeb\cdot\nabla_\k F(\k,\aeb)d\aeb\nonumber\\
& &=\i(\nabla_{\k}\cdot\nabla_{\x})\w_{kx}\x\cdot\nabla_{\x}\at_{\k\x}+\i(\nabla_{\k}\cdot\nabla_{\x})\w_{\k\x}\at_{\k\x}+
(\nabla_{\k}\cdot\nabla_{\x})\w_{\k\x}(\nabla_{\k}\cdot\nabla_{\x})\at_{\k\x}.\label{f1F1a0}
\EEA
\section{Calculation of $\xi$}
\label{app_xi}
In order to calculate $\xi$, we solve Eq.~(\ref{coshu}).
Notice that $\cosh 2\xi >1$ since $\mu_{ev}>\nu$.
Using the definition of the $\cosh$ function, we can write
\BEA
e^{2\xi}+e^{-2\xi}=2\frac{\mu_{ev}}{\sqrt{\mu_{ev}^2-\nu^2}},
\nn\EEA
and denoting $t=e^{2\xi}$, one obtains a quadratic equation for $t$ with two solutions
\BEA
t=\frac{\mu_{ev}\pm|\nu|}{\sqrt{\mu_{ev}^2-\nu^2}}.
\nn\EEA
After we take into account Eq.~(\ref{sinhv}), we obtain the following expression for $\xi$
\BEA
 \xi = \frac{1}{4}\ln{\frac{\muev-\nu}{\muev+\nu}}.\label{ksi_def}
\EEA
\section{Bogolyubov transformation of the $O(\ep)$ part}
\label{app_Bogolyubov}
Here, we show how Bogolyubov transformation works on $H_{f,\ep}^{(1)}$.
We consider the terms of Eq.~(\ref{H1e}) starting with
\BEA
&-&\int\left[\ac(\x\cdot\nabla_\x\mu)\ac^*+\frac{1}{2}(\x\cdot\nabla_\x\nu)(\ac\ac_-+\ac^*\ac^*_-)\right]d\k d\x\nonumber\\
=&-&\int(ub +vb_-^*)(\x\cdot\nabla_\x\mu)(ub^*+vb_-)d\k d\x+\left[\frac{1}{2}\int(\x\cdot\nabla_\x\nu)(u b +  v b_-^*)(u b_- +v b^*) d\k d\x +c.c.\right]\nonumber\\
=&-&\int b \left[ (u^2+v^2)(\x\cdot\nabla_\x\muev)+(u^2 - v^2)(\x\cdot\nabla_\x\muod) + 2uv(\x\cdot\nabla_\x\nu)\right]b^*d\k d\x-\nonumber \\
& &~~~-\int(u v (\x\cdot\nabla_{\x} \mu)+\frac{1}{2}(u^2+v^2) (\x\cdot \nabla_\x\nu))(b b_- +b^*b_-^*)d\k d\x\nonumber\\
=&&\int\left[-b(\x\cdot\nabla_{\x} \omega)b^*+\frac{\muev^2}{2\nu}\left(\x\cdot\nabla_\x\sqrt{1-\frac{\nu^2}{\muev^2}}\right)
\left(bb_- + b^* b_-^*\right)\right]d\k d\x.\label{O1e1}
\EEA
Here we used the following equalities
\BEA
& &u^2+v^2=\cosh(2\xi),\ \ 2uv=\sinh(2\xi),\ \ u^2-v^2=1,\nonumber\\
& &\sinh(2\xi)=\partial_{\nu}\omega,\ \ \cosh(2\xi)=\partial_{\muev}\omega,\ \ 1=\partial_{\muod}\omega,\nonumber\\
&
&\sinh(2\xi)\nabla_\x\muev+\cosh(2\xi)\nabla_\x\nu=-\frac{\muev^2}{\nu}\nabla_\x\sqrt{1-\frac{\nu^2}{\muev^2}}.\label{property}
\EEA
Next, we consider the terms of Eq.~(\ref{H1e}) with the Poisson bracket
\BEA
& &\int\i\ac\{\mu,\ac^*\}d\k d\x+\frac{1}{2}\int\i\{\nu,\ac_-\}d\k d\x=\frac{\i}{2}\int\Big(\ac\{\mu,\ac^*\}+\ac\{\nu,\ac_-\}\Big)d\k d\x +c.c.\nonumber\\
& &=\frac{\i}{2}\int(u b + v b_-^*)\Big(\{\mu,u b^* + v b_-\}+\{\nu,u b_- + v b^*\}\Big)d\k d\x+c.c.\label{rest}
\EEA
Note that\begin{itemize}
    \item $bb^*$ terms give zero because their coefficients are purely imaginary and we add c.c. values in the end,
    \item $b^*\nabla b$ terms can be obtained as c.c. of $b\nabla b^*$ terms.
\end{itemize}
Here, $\nabla$ denotes a gradient either with respect to $\x$ or with respect to $\k$.
Now, let us consider (\ref{rest}) term by term.
\begin{enumerate}
    \item $b\nabla b^*$ and $b^*\nabla b$:\\
    \BEA
    & &\frac{\i}{2}\int \left[u^2b\{\mu,b^*\}+\underbrace{v^2b_-^*\{\mu,b_-\}+vub_-^*\{\nu,b_-\}}_
    {\mbox{change $\k\rightarrow-\k$}}
    +uvb\{\nu,b^*\}\right]d\x d\k+c.c.\nonumber \\
    & &=\frac{\i}{2}\int \Big(b\{\w,b^*\}-b^*\{\w,b\}\Big)d\x d\k=\i\int b\{\omega,b^*\}d\x d\k.\label{first}
    \EEA
    Here, we used the fact that $\int b\{\w,b^*\}d\k d\x=-\int b^*\{\w,b\}d\k d\x$
    \item $bb_-$ and $b^*b_-^*$:\\
    \BEA
    & &\frac{\i}{2}\int\left[ubb_-\{\mu,v\}+vb_-^*b^*\{\mu,u\}+\underbrace{ubb_-\{\nu,u\}+vb_-^*b^*\{\nu,v\}}_
    {\mbox{give 0 because $\nu$ is even}}\right]
    d\k d\x+c.c.\nonumber \\
    & &=\frac{\i}{2}\int\Big[(bb_--b^*b_-^*)(u\{\mu,v\}-v\{\mu,u\})\Big]d\k d\x
    =\frac{i}{2}\int\left[(bb_--b^*b_-^*)\underbrace{\{\muev+\muod,\xi\}}_{\mbox{only $\muod$ survives}}\right]d\k d\x\nonumber\\
    & &=\frac{\i}{2}\int\Big[(bb_--b^*b_-^*)\{\muod,\xi\}\Big]d\k d\x.\label{second}
    \EEA
    Here, we have used Eq.~(\ref{ksi_def}).
    \item $b\nabla b_-$ and $b^*\nabla b_-^*$:\\
    \BEA
    & &\frac{\i}{2}\int\Big[uvb_-^*\{\mu,b^*\}+uvb\{\mu,b_-\}+u^2b\{\nu,b_-\}+v^2b_-^*\{\nu,b^*\}\Big]
    d\k d\x+c.c.\nonumber \\
    & &=\frac{\i}{2}\int\Big[b\sinh(2\xi)\{\muev,b_-\}+b\cosh(2\xi)\{\nu,b_-\}\Big]d\k d\x+c.c.\nonumber \\
    & &=\frac{\i}{2}\int b \left[\frac{\muev^2}{\nu}\left\{\sqrt{1-\frac{\nu^2}{\muev^2}},b_-\right\}\right]d\k d\x+c.c.\label{third}
    \EEA
    We used Eq.~(\ref{property}) here.
\end{enumerate}
Finally, the rest of $O(\varepsilon)$ terms are
\BEA
\frac{\i}{2}\int\tilde{\nu}aa_-d\k d\x+c.c.=\frac{\i}{2}\int\tilde{\nu}bb_-d\k d\x+c.c.\label{nutilda}
\EEA
Combining Eqs.~(\ref{O1e1}), (\ref{first}), (\ref{second}), (\ref{third}), and~(\ref{nutilda}), we obtain the $O(\ep)$ part of the Hamiltonian given by
Eq.~(\ref{HafterZLF}).
\section{Canonicity conditions for near-identity transformation}
\label{app_canonicity}
Here, we obtain the canonicity conditions for the coefficients of the near-identity transformation~(\ref{nit}).
For the canonicity up to $O(\varepsilon)$ order, we use the equation of motion in the Hamiltonian form.
\BEA
\i\dot{b}_\k=\frac{\delta H}{\delta b_\k^*}.\label{motion}
\EEA
In this Appendix for simplicity of notation, we skip writing $\x$ in the subscript of the dynamical variables.
Using Eq.~(\ref{nit}), we obtain
\BEA
\i\Big(\dot{c}_\k+\alk\dot{c}_{-\k}^*+\bk\{\gk,\dot{c}_{-\k}^*\}\Big)=\frac{\delta H}{\delta b_\k^*}.\label{motionchanged}
\EEA
Since we are neglecting terms of the order higher than $O(\ep)$, we can use the following approximation
\BEA
\dot{c}_{-\k}^*=\i\frac{\delta H}{\delta c_{-\k}}\approx \i\omega_{-\k}c_{-\k}^*.\label{ckapprox}
\EEA
Combining Eqs.~(\ref{motionchanged}) and~(\ref{ckapprox}), we obtain
\BEA
\frac{\delta H}{\delta b_\k^*}=\frac{\delta H}{\delta c_\k^*}-\alk\w_{-\k}c_{-\k}^*-\bk\{\gk,\w_{-\k}c_{-\k}^*\}.\label{bcherezc}
\EEA
Further, we have the chain rule in the form
\BEA
\frac{\delta H}{\delta c_\k^*}=\int\left(\frac{\delta H}{\delta b_\q^*}\frac{\delta b_\q^*}{\delta c_\k^*}+\frac{\delta H}{\delta b_{-\q}}
\frac{\delta b_{-\q}}{\delta c_\k^*}\right)d\q.\label{integral}
\EEA
Using Eq.~(\ref{nit}), we find
\BEA
\frac{\delta b_\q^*}{\delta c_\k^*}&=&\delta_\k^\q,\nonumber \\
\frac{\delta b_{-\q}}{\delta c_\k^*}&=&\alpha_{-\q}\delta_\k^\q-\beta_{-\q}\{\gamma_{-\q},\delta_\k^\q\}_\q,\nonumber
\EEA
where the subscript of the Poisson bracket indicates the differentiation with respect to $\q$.
Therefore, Eq.~(\ref{integral}) becomes
\BEA
\frac{\delta H}{\delta c_\k^*}=\frac{\delta H}{\delta b_\k^*}+\alpha_{-\k}\omega_{-\k}c_{-\k}^*+\{\gamma_{-\k},\beta_{-\k}\w_{-\k}c_{-\k}^*\}_\k.\label{ccherezb}
\EEA
Combining Eqs.~(\ref{bcherezc}) and~(\ref{ccherezb}), we find
\BEA
0=-2\w_{-\k}c_{-\k}^*\alod-\bk\{\gk,\w_{-\k}c_{-\k}^*\}+\{\gmk,\bmk\w_{-\k}c_{-\k}^*\}.\label{canon0}
\EEA
Finally, we obtain the canonicity conditions given in Eq.~(\ref{cancon}).
\section{Near-identity transformation}
\label{app_nit}
Lets us first apply the near-identity transformation to the $O(1)$ part of the Hamiltonian that is given by Eq.~(\ref{H1afterBogolyubov})
\BEA
\int\w b^*b~d\k d\x=\int\w c^*c+\big(\w\al^*cc_-+\w\b^*c\{\g^*,c_-\}+c.c.\big)d\k d\x+h.o.t.
\nn\EEA
To apply this transformation to the $O(\varepsilon)$ part we just need to substitute $b$ with $c$ in Eq.~(\ref{HeafterZLF}).
The non-diagonal terms cancel in the Hamiltonian if
\BEA
\int\big(\sigma+\w\al^*\big)cc_-+\left(\w\b^*c\{\gamma^*,c_-\}+\frac{\i\muev^2}{2\nu}c\{\varphi,c_-\}\right)d\k d\x=0.\label{diagcon1}
\EEA
Let us choose
\BEA
\g=\g^*=\varphi=\frac{\w_{ev}}{\mu_{ev}}.\label{gammacon}
\EEA
Then we can rewrite Eq.~(\ref{diagcon1}) as
\BEA
\int(\sigma+\omega\al^*)cc_-+\left((\wev+\wod)\beta^*+\frac{\i\muev^2}{2\nu}\right)c\{\varphi,c_-\}d\k d\x=0.
\nn\EEA
Integrating by parts one can show that
\BEA
\int\wod\beta^*c\{\varphi,c_-\}d\k d\x=\frac{1}{2}\int\{\wod\beta^*,\varphi\}cc_-d\k d\x.
\nn\EEA
Therefore, the diagonalizing condition becomes
\BEA
\int\left(\sigma+\omega\al^*+\frac{1}{2}\{\wod\beta^*,\varphi\}\right)cc_-+\left(\wev\beta^*+\frac{\i\muev^2}{2\nu}\right)c\{\varphi,c_-\}d\k d\x=0\label{diagcon2}
\nn\EEA
From Eq.~(\ref{diagcon2}), the condition on $\beta$ immediately follows
\BEA
\b=\frac{\i\muev^2}{2\nu\wev}.\label{app_beta}
\EEA
In order to obtain the condition on $\al$, we expand the second term in the integral
\BEA
\w\al^*=\wod\alod^*+\wev\alev^*+\wod\alev^*+\wev\alod^*.\label{expand}
\EEA
Integral over the last two terms vanishes because these functions are odd.
Therefore, we consider only the other two terms.
Next, we insert this expansion into Eq.~(\ref{diagcon2})
\BEA
\int\left(\sigma+\wod\alod^*+\wev\alev^*+\frac{1}{2}\wod\{\beta^*,\varphi\}+\frac{1}{2}\beta^*\{\wod,\varphi\}\right)cc_-
+\left(\wev\beta^*+\frac{i\muev^2}{2\nu}\right)\{\varphi,c_-\}cdkdx=0.
\nn\EEA
Then, we obtain the following diagonalizing conditions on $\alpha$
\BEA
\alpha_{ev}&=&-\frac{\sigma^*}{\wev}-\frac{\b}{2\wev}\left\{\wod,\frac{\w_{ev}}{\mu_{ev}}\right\},\label{app_alpha_ev}\\
\alpha_{od}&=&-\frac{1}{2}\left\{\b,\frac{\wev}{\mu_{ev}}\right\}.\label{app_alpha_od}
\EEA
Let us prove that $\al_{od}=0$.
Substituting Eq.~(\ref{app_beta}) into Eq.~(\ref{app_alpha_od}), we obtain
\BEA
\al_{od}=-\frac{i}{4}\left\{\frac{\muev^2}{\nu\wev},\frac{\omega_{ev}}{\mu_{ev}}\right\}.\nonumber
\EEA
Expanding the Poisson bracket we find the following identity
\BEA
\left\{\frac{\muev^2}{\nu\wev},\frac{\omega_{ev}}{\mu_{ev}}\right\}=
\frac{\nu\{\mu_{ev},\w_{ev}\}+\mu\{\w_{ev},\nu\}+\w_{ev}\{\nu,\mu_{ev}\}}{\nu^2\omega_{ev}}.\label{A4_bracket}
\EEA
According to the definition, $\w_{ev}^2=\mu_{ev}^2-\nu^2$.
Differentiation of both sides of the last equality with respect to $\x$ or $\k$ yields
\BEA
\nabla\w_{ev}=\frac{\mu_{ev}}{\w_{ev}}\nabla\mu_{ev}-\frac{\nu}{\w_{ev}}\nabla\nu.\label{app_w_prime}
\EEA
We use Eq.~(\ref{app_w_prime}) to show that
\BEA
\{\mu_{ev},\w_{ev}\}&=&-\frac{\nu}{\w_{ev}}\{\mu_{ev},\nu\},\nonumber\\
\{\w_{ev},\nu\}&=&\frac{\mu_{ev}}{\w_{ev}}\{\mu_{ev},\nu\}.\nonumber\\
\label{app_tmp}
\EEA
Plugging Eq.~(\ref{app_tmp}) into Eq.~(\ref{A4_bracket}) proves that $\al_{od}=0$.


\begin{thebibliography}{10}
\bibitem{Arnold} V.I. Arnold, Mathematical Methods of Classical Mechanics, Springer-Verlag, 1978.
\bibitem{Zakharov_grav} A. Pushkarev, D. Resio, V. Zakharov,
``Weak turbulent approach to the wind-generated gravity sea waves'', Physica D 184 (1-4) 29-63, 2003.
\bibitem{monin_piterbarg} A.S. Monin, L.I. Piterbarg, ``On the kinetic equation of Rossby-Blinova waves'',
Doklady Akademii Nauk SSSR 295 (4), 816-820, 1987.

\bibitem{Balk_naz_zakh} A.M. Balk, V.E. Zakharov and S.V. Nazarenko,
``Nonlocal Drift Wave Turbulence'',
 Sov.Phys.-JETP 71, 249-260, 1990.
\bibitem{Lvov_Tabak} Yu. Lvov, E. Tabak, ``Hamiltonian formalism and the Garrett-Munk spectrum of internal waves in the ocean''
PRL vol. 87, 168501,2001.
\bibitem{galtier_etal_2000} S. Galtier, S.V. Nazarenko, A.C. Newell,
A. Pouquet, ``A weak turbulence theory for incompressible MHD'' 
J. Plasma Phys., 63, 447, 2000.
\bibitem{BEC} V.E. Zakharov, S.V. Nazarenko, ``Dynamics of the Bose-Einstein condensation'', Physica D, 201, 203 (2005).
\bibitem{BEC_LN} Y.V. Lvov, S. Nazarenko and R. West, ``Wave turbulence in Bose–Einstein condensates'', Physica D 184, 333 2003.
\bibitem{Plasma} A.A. Galeev, R.Z. Sagdeev, in Reviews of Plasma
Physics", Vol. 6 (Ed. M.A. Leontovich) New York: Consultants Bureau,
1973.
\bibitem{ZLF} V.E. Zakharov, V.S. Lvov, G. Falkovich, "Kolmogorov Spectra of Turbulence", Springer-Verlag, 1992.
\bibitem{Capil} A.N. Pushkarev, V.E. Zakharov, ``Turbulence of capillary waves — theory and numerical simulation'', Physica D 135 (1-2) 98-116, 2000.
\bibitem{Aged}V.E. Zakharov, ``Direct and inverse cascade in wind-driven sea and wave breaking'', Proceedings of IUTAM Meeting on Wave Breaking (Sydney, 1991),
eds. M.L. Banner and R.H.Y. Grimshaw (Springer-Verlag, Berlin, 1992), pp. 69-91
\bibitem{W} Gregor Wentzel ``Eine Verallgemeinerun der
Quantenbedingungen für die Zwecke der Wellenmechanik'', Z. Physik. 38
518-529 (1926).
\bibitem{K} H. A. Kramers ``Wellenmechanik und halbzahlige
Quantisierung'', Zeits. Physik. 39 828-840 (1926).
\bibitem{B} Léon Brillouin, ``La mécanique ondulatoire de
Schrödinger; une méthode générale de resolution par
approximations successives'', Comptes rendus (Paris) 183 24-26 (1926).
\bibitem{janssen} P. Janssen, "The interaction of ocean waves and wind", Cambridge University Press, 2004.
\bibitem{LandauStat} L.D. Landau and E.M. Lifshitz, Statistical
 Physics, (Course of Theoretical Physics, Vol 5), Pergamon Press, (1969).
\bibitem{LandauKin} E.M. Lifshitz and L.P. Pitaevskii,
Physical Kinetics (Course of Theoretical Physics, Vol 10).
Butterworth-Heinemann (1981)
\bibitem{mallat} S. Mallat, "A wavelet tour of signal processing", Academic Press, 1998.
\bibitem{VSLvovLectures} V.S.Lvov and A. M. Rubenchik, ``Spatially
non-uniform singular weak turbulence spectra'',Sov. Phys. JETP 45 (10)
pp. 67 - 74 (1977).\\  Also at V.S. Lvov, ``Fundamental of Nonlinear
Physics, Lecture Notes'', Equation (7.8), www.lvov.weizmann.ac.il,
(2005).
\bibitem{Bose} S. N. Bose. "Plancks Gesetz und Lichtquantenhypothese",
Zeitschrift für Physik 26:178-181 (1924).


\bibitem{Einst1} A. Einstein, Sitzber. Kgl. Preuss. Akad Wiss., 261
(1924).  Albert Einstein, ``Quantentheorie des einatomigen idealen
Gases / Quantum Theory of ideal Monoatomic Gases'',
Sitz. Ber. Preuss. Akad. Wiss. (Berlin) 22, 261 (1924);\\ Reprinted in
The collected papers of Albert Einstein, ed. by J. Stachel, Princeton
University Press (1989).
\bibitem{Einst2} 
Albert Einstein,
``Quantentheorie des einatomigen idealen Gases / Quantum Theory of ideal
Monoatomic Gases''
Sitz. Ber. Preuss. Akad. Wiss. (Berlin) 23, 3 (1925);\\
Reprinted in
The collected papers of Albert Einstein, ed. by J. Stachel, Princeton
University Press, (1989).
\bibitem{Anderson} M.H. Anderson, et al., ``Observation of Bose-Einstein Condensation in a Dilute Atomic Vapor'', 
Science 269, 198 (1995).
\bibitem{Bradley} C.C. Bradley, et al., ``Evidence of Bose-Einstein
Condensation in an Atomic Gas with Attractive Interactions'',
Phys. Rev. Lett. 75, 1687 (1995).
\bibitem{Davis} K.B. Davis, et al., ``Bose-Einstein Condensation in a Gas of Sodium Atoms''
 Phys. Rev. Lett. 75, 3969 (1995).
\bibitem{Pitaebsky} 
L. P. Pitaevskii, È Vortex lines in an imperfect Bose gas,É Zh. Eksp. Teor.
Fiz. 40, 646 (1961) [Sov. Phys. JETP 13, 451 (1961)].
\bibitem{dnz} A.I. Dyachenko, S.V. Nazarenko and V.E. Zakharov, 'Wave-vortex              dynamics in drift and $\beta$-plance turbulence. 
Phys,Lett.A 165,  330-334 (1992).
\bibitem{rudak} A.A. Vedenov, A.V. Gordeev and L.I. Rudakov, 
``Oscillations and instability of a weakly turbulent plasma''
Plasma Physics 9, 719 (1967).
\end{thebibliography}
\end{document}